\documentclass[10pt,conference]{IEEEtran}

\newif\ifcameraready
\camerareadytrue
\newif\ifarxiv
\arxivfalse  
\ifdefined\arxivbuild\arxivtrue\fi
\newcommand{\arxivonly}[1]{\ifarxiv#1\fi}
\newif\ifpublicurls
\ifcameraready\publicurlstrue\fi
\ifarxiv\publicurlstrue\fi
\newcommand{\dataseturl}{\ifpublicurls\url{https://huggingface.co/datasets/cezar-sas/code-theseus}\else\url{https://anonymous-hf.com/a/0c9u63n13mvu/}\fi}
\newcommand{\codeurl}{\ifpublicurls\url{https://github.com/SasCezar/code-theseus}\else\url{https://anonymous.4open.science/r/code-theseus/}\fi}

\usepackage[numbers,sort&compress]{natbib}
\usepackage{amssymb}
\usepackage{graphicx}
\usepackage{booktabs}
\usepackage{textcomp}
\usepackage[hidelinks]{hyperref}

\usepackage[inline]{enumitem}

\usepackage{xspace}
\newcommand{\README}{\texttt{README}\xspace}
\newcommand{\cIntent}{\textsc{Intent}\xspace}
\newcommand{\cFull}{\textsc{Full}\xspace}
\newcommand{\mGemma}{GEMA\xspace}
\newcommand{\mQwen}{QW35\xspace}
\newcommand{\mGpto}{GPTO\xspace}
\newcommand{\ie}{i.e.,\@\xspace}
\newcommand{\eg}{e.g.,\@\xspace}
\usepackage{tablefootnote}
\usepackage{multirow}

\newcommand{\upgood}[1]{\textcolor{better}{\scriptsize $\blacktriangle$\,#1}}
\newcommand{\upbad}[1]{\textcolor{worse}{\scriptsize $\vartriangle$\,#1}}
\newcommand{\downgood}[1]{\textcolor{better}{\scriptsize $\blacktriangledown$\,#1}}
\newcommand{\downbad}[1]{\textcolor{worse}{\scriptsize $\triangledown$\,#1}}
\xspaceaddexceptions{]\%}
\newcommand{\AgentLibs}{29\xspace}
\newcommand{\AgentRuns}{340\xspace}

\newcommand{\AgentRunsExcluded}{8\xspace}

\newcommand{\HeldOutReplicateSD}{12.0\xspace}
\newcommand{\AgentMaxSteps}{1,000\xspace}
\newcommand{\AtCapPct}{69\xspace}
\newcommand{\AtCapIntent}{68\xspace}
\newcommand{\AtCapFull}{70\xspace}

\newcommand{\HeldOutIntent}{33.7\xspace}
\newcommand{\HeldOutFull}{35.8\xspace}
\newcommand{\HeldOutDiff}{\ensuremath{+}2.1\xspace}
\newcommand{\HeldOutCILow}{\ensuremath{-}1.6\xspace}
\newcommand{\HeldOutCIHigh}{\ensuremath{+}5.8\xspace}

\newcommand{\HeldOutHolm}{0.160\xspace}

\newcommand{\ModFLibs}{24\xspace}
\newcommand{\ModFIntent}{0.54\xspace}
\newcommand{\ModFFull}{0.63\xspace}
\newcommand{\ModFDiff}{\ensuremath{+}0.09\xspace}
\newcommand{\ModFCILow}{\ensuremath{+}0.05\xspace}
\newcommand{\ModFCIHigh}{\ensuremath{+}0.15\xspace}
\newcommand{\ModFHigher}{17\xspace}

\newcommand{\ModFHolm}{0.005\xspace}

\newcommand{\ModuleRecallDiff}{\ensuremath{+}0.12\xspace}

\newcommand{\AllModFIntent}{0.77\xspace}

\newcommand{\AriHolm}{0.160\xspace}

\newcommand{\TaskTokensIntent}{1.0\xspace}
\newcommand{\TaskTokensFull}{4.8\xspace}

\newcommand{\AgentStepsIntent}{869\xspace}
\newcommand{\AgentStepsFull}{858\xspace}

\newcommand{\DocRecallIntent}{0.58\xspace}
\newcommand{\DocRecallFull}{0.70\xspace}

\newcommand{\VisiblePassRateIntent}{49.8\xspace}
\newcommand{\VisiblePassRateFull}{51.2\xspace}

\newcommand{\HeldOutMaxGain}{\ensuremath{+}24.3\xspace}

\newcommand{\HeldOutMaxLoss}{\ensuremath{-}21.9\xspace}

\newcommand{\ContractCovMedian}{67\xspace}
\newcommand{\ContractCovMin}{7\xspace}
\newcommand{\ContractCovMax}{100\xspace}
\newcommand{\ContractFullLibs}{5\xspace}

\newcommand{\ModFCovRho}{\ensuremath{+}0.29\xspace}
\newcommand{\ModFCovP}{0.162\xspace}

\newcommand{\ModFDocCovRho}{\ensuremath{+}0.07\xspace}
\newcommand{\ModFDocCovP}{0.756\xspace}
\newcommand{\ModFInteractionP}{0.099\xspace}

\newcommand{\ModFLowCovDiff}{\ensuremath{+}0.05\xspace}

\newcommand{\ModFHighCovDiff}{\ensuremath{+}0.15\xspace}

\newcommand{\HeldOutCovP}{0.214\xspace}

\newcommand{\HeldOutInteractionP}{0.230\xspace}

\newcommand{\HeldOutLowCovDiff}{\ensuremath{+}3.1\xspace}

\newcommand{\HeldOutHighCovDiff}{\ensuremath{+}1.2\xspace}

\newcommand{\ModFGenInterP}{0.484\xspace}

\newcommand{\CovGainUnits}{48\xspace}

\newcommand{\DocCovSD}{32\xspace}
\newcommand{\LengthSD}{1.7\xspace}
\newcommand{\ModFCovGainBeta}{\ensuremath{+}0.011\xspace}
\newcommand{\ModFCovGainP}{0.664\xspace}
\newcommand{\ModFLenGainBeta}{\ensuremath{+}0.021\xspace}
\newcommand{\ModFLenGainP}{0.315\xspace}
\newcommand{\ModFCovGainAdjBeta}{\ensuremath{+}0.004\xspace}
\newcommand{\ModFCovGainAdjP}{0.884\xspace}

\newcommand{\HeldOutLenGainBeta}{\ensuremath{-}3.5\xspace}
\newcommand{\HeldOutLenGainP}{0.125\xspace}

\newcommand{\GemmaUnitsNamed}{13\xspace}

\newcommand{\GemmaDiagramsValid}{94.3\xspace}
\newcommand{\GemmaDocsValid}{82.4\xspace}

\newcommand{\QwenUnitsNamed}{29\xspace}

\newcommand{\QwenDiagramsValid}{83.8\xspace}
\newcommand{\QwenDocsValid}{53.4\xspace}

\newcommand{\GptoUnitsNamed}{31\xspace}

\newcommand{\GptoDiagramsValid}{57.8\xspace}
\newcommand{\GptoDocsValid}{20.0\xspace}
\newcommand{\DocsTotal}{9,000\xspace}

\newcommand{\DiagramsFailed}{3,666\xspace}
\newcommand{\FailParse}{16\xspace}
\newcommand{\FailUndefined}{83\xspace}

\newcommand{\UndefinedElsewhere}{69\xspace}

\newcommand{\GemmaQ}{3.48\xspace}

\newcommand{\QwenQ}{3.02\xspace}

\newcommand{\GptoQ}{2.84\xspace}

\newcommand{\HolisticRhoMean}{0.89\xspace}

\newcommand{\QIccSingle}{0.86\xspace}
\newcommand{\QIccMean}{0.95\xspace}

\newcommand{\SizeRequirementsRhoMean}{-0.18\xspace}
\newcommand{\SizeArchitectureRhoMin}{-0.24\xspace}

\newcommand{\PopularityRhoAbsMax}{0.07\xspace}

\newcommand{\PopularityIntentRhoMax}{+0.12\xspace}

\newcommand{\AlignIntentCovMin}{63\xspace}
\newcommand{\AlignIntentCovMax}{72\xspace}
\newcommand{\AlignIntentCovShuffled}{5\xspace}
\newcommand{\AlignArchCovMin}{78\xspace}
\newcommand{\AlignArchCovMax}{87\xspace}
\newcommand{\AlignArchCovShuffled}{10\xspace}

\usepackage[most]{tcolorbox}
\usepackage{xcolor}
\usepackage{xparse}
\usepackage{amsmath}

\definecolor{accentgreen}{HTML}{4E9A59}
\definecolor{better}{HTML}{1A6B2F} 
\definecolor{worse}{HTML}{C0622B}  
\definecolor{framegray}{HTML}{D9D9D9}
\definecolor{pillbg}{HTML}{E8F5EA}

\newtcolorbox[
  auto counter,
  number within=section
]{prompt}[2][]{%
  enhanced,breakable,
  colback=white,colframe=white,boxrule=0.6pt,arc=0pt,
  left=10pt,right=10pt,top=10pt,bottom=10pt,
  borderline west={3pt}{0pt}{accentgreen},
  attach boxed title to top left={yshift*=-\tcboxedtitleheight/2},
  boxed title style={
    colback=pillbg,colframe=accentgreen,boxrule=0.6pt,arc=0pt,
    left=8pt,right=8pt,top=4pt,bottom=4pt
  },
  title=\textbf{\color{accentgreen}Prompt~\thetcbcounter: #2},
  #1
}

\definecolor{takeawaydark}{HTML}{2F6B3A}
\newtcolorbox[auto counter]{takeaway}[2][]{%
  enhanced,
  colback=pillbg,colframe=pillbg,boxrule=0pt,arc=0pt,
  borderline west={2pt}{0pt}{takeawaydark},
  left=6pt,right=6pt,top=4pt,bottom=4pt,
  before upper={\tcbox[on line,colback=takeawaydark,colframe=takeawaydark,
      coltext=white,boxrule=0pt,arc=1.5pt,boxsep=0pt,left=2.5pt,right=2.5pt,
      top=1pt,bottom=1pt,fontupper=\scriptsize\bfseries]{TAKEAWAY~\thetcbcounter}%
    \enspace\textbf{\color{takeawaydark!85!black}#2.}\ },
  fontupper=\small,#1
}

\begin{document}

\bstctlcite{IEEEexample:BSTcontrol}

\title{The Code of Theseus: Can Agents Rebuild Software from Recovered Design Documents?}

\ifcameraready
\author{
\IEEEauthorblockN{
    Cezar Sas\IEEEauthorrefmark{1},
    Sandra Mitrovi\'{c}\IEEEauthorrefmark{1},
    Fabio Barth\IEEEauthorrefmark{2},
    Georg Rehm\IEEEauthorrefmark{2},
    Matteo Salani\IEEEauthorrefmark{1}
}

\IEEEauthorblockA{
    \IEEEauthorrefmark{1}IDSIA (SUPSI), Switzerland\\
    \{cezar.sas, sandra.mitrovic, matteo.salani\}@idsia.ch
}

\IEEEauthorblockA{
    \IEEEauthorrefmark{2}DFKI, Germany\\
    \{fabio.barth, georg.rehm\}@dfki.de
}
}
\else
\author{\IEEEauthorblockN{Anonymous Author(s)}}
\fi

\maketitle

\begin{abstract}
LLM agents increasingly design a system before implementing it, yet the data to train and evaluate them centres on implementation and testing: requirements and design artefacts are often missing, outdated or private. Source code, in contrast, is abundant. Prior work recovers architecture documentation by feeding LLMs the source code or letting agents explore it, which saturates the context, does not scale to many repositories, and yields a single artefact rather than aligned ones.
We present a framework that recovers aligned stakeholder intent, requirements and architecture from a compact repository representation, a static dependency graph and key contextual files, and release the artefacts of 10,764 Java, Python, C/C++ and C\# repositories as a dataset. An LLM judge rates the artefacts of three generators as clear and internally consistent, but less reliably correct and at the right level of abstraction.
To test whether the artefacts carry the design, we pose a software Ship of Theseus: a coding agent rebuilds 29 \textsc{Commit0} libraries from the intent alone or with the recovered requirements and architecture, given only an interface contract and part of the tests. The artefacts do not significantly change the held-out pass rate (\HeldOutIntent\% vs.\ \HeldOutFull\%), but bring the rebuilt module structure significantly closer to the original (module F1 from \ModFIntent to \ModFFull), with either generator and even with concise artefacts.
Recovered artefacts thus convey design that tests do not capture, offering a scalable source of design-level data for agents that design before they implement, and the round trip measures this without human references.
\end{abstract}


\section{Introduction}
Large language model (LLM) agents are increasingly used to automate software engineering (SE) tasks, from code generation and testing to debugging and repository-level maintenance. Multi-agent frameworks such as MetaGPT~\cite{hong2024metagpt}, AgileCoder~\cite{nguyen2025agilecoder}, HyperAgent~\cite{phan2024hyperagent} and ChatDev~\cite{qian2024chatdev} go further and distribute the work across the stages of the software development lifecycle (SDLC), so that the system is designed before it is implemented.

The data available to train and evaluate such agents does not follow this shift. Existing benchmarks focus on implementation (61\%) and testing (12\%), while requirements engineering (5\%) and software design (3\%) are barely represented~\cite{wang2025bench_sdlc}. Yet these early stages are necessary before implementation, whether the agent plans its own work or a multi-agent system assigns requirements and design to dedicated roles. Without them, agents write monolithic programs, with far fewer files and shallower directories than the originals~\cite{yang2026programbench}. Data that connects intent, requirements and design to the code that realises them is scarce, because early-stage artefacts are often missing, incomplete, obsolete, inconsistent with the implementation, or private~\cite{pauzi2025traceability, xu2025autoreqgen}. This also poses a limitation to comprehensive benchmarking of multi-agent systems for software development on intermediate steps~\cite{mitrovic2026eumas}.

Source code, in contrast, is abundant, and LLMs can recover design documentation from it. CIAO~\cite{deluca2026ciao} reconstructs architecture documentation that contributors to the repository find useful, from a flattened representation of its source code and directory structure. Two questions remain open if recovered artefacts are to serve as data for agents. First, generation must scale to thousands of repositories and produce artefacts that are aligned with each other, not architecture documentation alone. Second, it is unclear whether the artefacts carry enough of the design to build the system from them.

We address both. Our framework builds a compact, deterministic representation of each repository by using static analysis to extract a dependency graph and contextual files, which are then used to condition an LLM to generate three aligned artefacts: stakeholder intent, requirements, and architecture. Applied at scale, the framework yields aligned intent–requirements–architecture–code data on which LLMs can be trained.

We evaluate the artefacts in two ways, which give our three research questions. First, an LLM judge rates the artefacts of three generator models on seven quality dimensions (\textbf{RQ1}: \emph{how good are the recovered artefacts as a coherent body of documentation?}). Second, we close the loop from code to artefacts and back to code: a coding agent rebuilds 29 Python libraries of the \textsc{Commit0} benchmark~\cite{zhao2025commit0} from the artefacts of two generators alone, and the original implementation serves as the reference. This poses a software version of the Ship of Theseus: if an agent rebuilds a library from the design recovered from its code, is it the same system? The held-out tests show whether it behaves the same (\textbf{RQ2}), and structural metrics show whether it is built the same way (\textbf{RQ3}).

In summary, this paper makes four contributions:
\begin{itemize}[noitemsep]
    \item A framework for recovering aligned stakeholder intent, requirements, and architecture artefacts from compact repository representation;

    \item A round-trip evaluation showing improved module-level structural similarity without significant changes in functional correctness;

    \item An LLM-as-a-judge evaluation of three generator models across seven quality dimensions;

    \item A large-scale dataset of 10k repositories paired with aligned early-stage SDLC artefacts for LLM training.

\end{itemize}

The supplementary material are publicly available at \codeurl{} and as a Hugging Face dataset at \dataseturl.

\section{Related Work}
This work relates to several areas in AI-assisted SE and repository-level code understanding. We organise the related work into four areas: similar existing datasets, architecture recovery, requirements extraction, and software summarisation. Together, these areas explore how higher-level software-engineering artefacts can be derived from repositories.

\subsection{Existing Datasets}
Current datasets for AI-assisted software engineering primarily support implementation-oriented tasks, such as code generation, bug fixing, and repository-level maintenance, while comparatively few resources target earlier SDLC stages, including requirements engineering and software architecture design \cite{wang2025bench_sdlc}. More importantly, existing datasets rarely provide aligned artefacts spanning stakeholder intent, requirements, architecture, and implementation, limiting the development of agents capable of reasoning across the full software lifecycle.

Existing datasets can broadly be grouped into requirements-oriented, implementation-oriented, and architecture-related resources. For requirements engineering, \citet{shaukat2018dataset} published a dataset of software requirements specifications and associated risk attributes for requirements risk prediction. Although useful for classification, it neither provides holistic requirements elicitation nor connects requirements to architectural choices, and its data were collected by human experts rather than automatically generated. Other datasets contain domain-specific high- and low-level requirements \cite{10628478, 5328545, 6602474, 10.1145/2245276.2231943, 1583599}. PURE \cite{8049173}, for example, comprises 79 public requirements documents from diverse domains; however, these resources are generally not designed around architectural reasoning or broader SDLC processes, nor have they been adapted or annotated specifically for LLM training.

Implementation-oriented resources are substantially more common. Stack v2 \cite{lozhkov2024starcoder2stackv2} contains complete repositories alongside resolved user issues, software descriptions, and documentation, while SWE-Bench \cite{jimenez2024swebench} and SWE-Lancer \cite{miserendino2025swelancer} evaluate LLMs and agentic systems on repository-level code generation and maintenance tasks. \textsc{Commit0}~\cite{zhao2025commit0} and ProgramBench~\cite{yang2026programbench} ask agents to implement a library from its stubs and documentation, or to rebuild a program from its executable and documentation, and grade the result by tests. Related software summarisation and code-understanding datasets pair source code with natural-language artefacts such as comments, docstrings, commit messages, issue descriptions, and README files, ranging from function-level summaries to repository-level descriptions. These resources may implicitly capture purpose, intent, or requirements, but they target implementation, documentation, or code-to-text generation, not requirements elicitation or architectural design.

Architecture-focused datasets are even scarcer \cite{ESPOSITO2026112607}. LLM-based approaches to architectural design generation commonly rely on empirical studies or manually constructed examples rather than large-scale aligned datasets \cite{10592785, 10978918, 10628428}. For example, \citet{10.1145/3691620.3695019} published 185 design rationales derived from Jira issue logs. While these texts capture architectural design choices, they do not provide complete textual architecture descriptions or link those choices systematically to stakeholder intent, requirements, and implementation.

Overall, existing resources tend to address isolated SE tasks or individual SDLC stages rather than providing lifecycle-level alignment. To the best of our knowledge, no publicly available dataset systematically aligns stakeholder intent, software requirements, architectural descriptions, and source code. This gap motivates our reverse-engineering framework for generating aligned early-stage SDLC artefacts directly from existing software repositories.

\subsection{Software Architecture Recovery}
Software Architecture Recovery (SAR) reconstructs the architecture of a software system from available artefacts. Automated SAR techniques include clustering, pattern matching, and genetic-algorithm-based methods~\cite{qayum2025tertiaryms}. Among these, bottom-up graph-clustering approaches are the most widely used: they model relationships among software entities, such as files, classes, and functions, and group related entities into architectural modules.

Several approaches differ primarily in how they construct and enrich these dependency graphs. \citet{altinisik2024sar} construct separate graphs for call dependencies, evolutionary coupling, and module--database-table dependencies, cluster each using the Pivot algorithm, and aggregate the resulting partitions. SARIF~\cite{zhang23sarifusion} instead constructs a weighted dependency graph using Depends~\cite{depends2025}, with edge weights incorporating features such as Latent Dirichlet Allocation (LDA) derived source-code topics~\cite{blei03lda} and repository structure, before applying graph clustering. Because LDA-based similarity can lead to overly fragmented clusters~\cite{zhao2026semarc}, SemArc~\cite{zhao2026semarc} replaces it with semantic similarity, enriches the graph with implicit dependencies (\eg duck typing), and uses an LLM to identify architectural patterns. Despite these differences, all three approaches ultimately recover architectural components through graph partitioning.

More recently, SAR has shifted toward LLM-based recovery and documentation. \citet{amalfitano2025sarllm} investigate LLMs for identifying design and architectural patterns from source-code snippets, while \citet{pan2026archagent} use a multi-agent system to explore repositories, summarise files, and hierarchically reconstruct architectural representations. Related work uses retrieval-augmented LLMs to derive domain models from source code~\cite{boronat2025mdre}. At repository scale, CIAO~\cite{deluca2026ciao} generates human-readable system-level architecture documentation from a flattened representation of the source code and folder structure. It was reviewed on 22 small repositories, mostly by researchers who contributed to them. Other approaches pursue related goals: CodeWiki~\cite{codewiki2026} combines hierarchical decomposition with recursive multi-agent processing to generate architecture-aware documentation, while GitDiagram~\cite{khaleel2025gitdiagram} and DeepWiki~\cite{deepwiki2025} derive architectural descriptions for repository comprehension.

\subsection{Software Requirements Extraction}

Software requirements extraction recovers requirements from available software artefacts using automated or semi-automated techniques. Most prior work treats requirements analysis as an NLP-based classification problem over existing textual requirements~\cite{apte2023automatic}, whereas comparatively few studies infer intended functionality directly from source code~\cite{miskell2023automated}. Such recovered requirements can support program comprehension, maintenance, modernisation, and component reuse. We focus specifically on approaches that use NLP and LLMs to generate requirements-level artefacts, such as functional requirements and user stories, from source code~\cite{ouf2025reverse,xu2025autoreqgen}.

The central challenge is abstracting implementation details into externally meaningful system behaviour. Existing approaches generally operate on individual files, snippets, or small programs rather than complete repositories. They include hybrid static-analysis/LLM methods that extract structures such as ASTs and map them to functional requirements~\cite{miskell2023automated}; LLM-based generation of higher-level artefacts, such as user stories, from code inputs below 350 NLOC~\cite{ouf2025reverse}; and pipelines that select and organise evidence such as method definitions, variables, and comments before LLM-based generation and refinement~\cite{xu2025autoreqgen}. Thus, while these methods demonstrate that requirements-level information can be recovered from implementation artefacts, their scope remains narrower than repository-level reconstruction of a requirements specification.

\subsection{Software Summarization}

Although not our central focus, summarization is loosely related to some of our artefacts, so we briefly review it in the context of SE. Studies summarise bug reports~\cite{Chen_2021} and pull requests~\cite{Irsan_2022} to support bug identification and fixing, and code snippets~\cite{Sun_2025} and \README files~\cite{doan2023too} to facilitate repository comprehension. Automatic source code summarisation approaches can be categorised into template-based, information retrieval-based and deep learning-based~\cite{Zhang_review2024}. Recent studies exploit LLM-based approaches, investigating the effect of in-context learning and decoding strategies for code snippet summarization~\cite{Sun_2025} and introducing a BART- and T5-based approach to summarise lengthy \README files~\cite{doan2023too}. Moreover, \cite{Nguyen_2025} proposed Metagente, a teacher-student LLM-based multi-agent system for \README summarisation.

\section{Seed Repository Dataset}
\label{sec:seed_dataset}

We first build a seed dataset: the curated repositories and their representations, from which the artefacts are generated at scale (Section~\ref{sec:approach}). Its curation emphasises diversity, coverage, and quality, and comprises project selection, dependency graph construction, and repository representation, described below.

\subsection{Projects List}
To construct the initial list of repositories, we rely on a collection of various \textit{awesome-lists} repositories that curate frameworks and tools for specific domains—covering the programming languages of interest (Java, Python, C/C++, and C\#). We complement this set with projects that had already been analysed by Arcan~\cite{sas2023td_dataset}, ensuring the largest possible number of repositories compatible with our analysis pipeline.

For each candidate repository, we collect metadata, including the number of stars, the date of the most recent commit, and the archival status. We then apply filtering criteria based on these attributes to obtain a large, high-quality, and up-to-date subset. Specifically, we retain only repositories with at least one push in the last five years ($year \ge 2021$) and at least 50 stars. Of the 31,438 non-archived candidate repositories, this selection retains 14,525 (46\%\arxivonly{; see Appendix~\ref{app:list_filter}}).

\subsection{Dependency Graph}
We construct the dependency graph using a static analysis tool, rather than an LLM or agents as in previous work~\cite{pan2026archagent, deepwiki2025, khaleel2025gitdiagram}. While LLM-based approaches more naturally support multilingual programming projects, our design seeks to avoid issues such as context saturation, context-splitting and merging artefacts, hallucinations, non-deterministic behaviour, and ambiguity. These considerations align with the guidelines proposed by \citet{baltes2025guidelines} and the findings of \citet{amalfitano2025sarllm}. Furthermore, such approaches remain difficult to scale for large-scale dataset construction due to their substantial token and inference costs (Table~\ref{tab:token-compression}).

\begin{table}[h]
\centering
\caption{Source code vs.\ dependency graph token counts.}
\label{tab:token-compression}
\begin{tabular}{llrr}
\toprule
\textbf{Repository} & \textbf{Lang} & \textbf{Source} & \textbf{Dep.\ Graph (\% saved)} \\
\midrule
antlr/antlr4             & Java   & 653k & 56k\,(\textcolor{better}{\scriptsize $\blacktriangledown$\,\textbf{91.4\%}}) \\
microsoft/markitdown     & Python &  85k &  8k\,(\textcolor{better}{\scriptsize $\blacktriangledown$\,\textbf{90.6\%}}) \\
tesseract-ocr/tesseract  & C++    & 2.1M & 88k\,(\textcolor{better}{\scriptsize $\blacktriangledown$\,\textbf{95.8\%}}) \\
\bottomrule
\end{tabular}
\end{table}

For graph extraction, we use Arcan~\cite{sas2023arcan, fontana2017arcan}, a static analysis tool supporting Java, C/C++, C\#, and Python. Arcan produces a dependency graph representing software entities (\ie files, classes, packages, and components) as typed, labelled nodes and their dependencies as labelled, directed edges pointing from dependent entities to those they rely on. Arcan also extract \textit{units}, language-agnostic modular constructs (\eg packages, modules, or namespaces) annotated with identifiers, types, and structural metrics.

Arcan completes the analysis for 12,664 of the 14,525 selected repositories. Of these, we keep those whose dependency graph has between the 10th and 95th percentile of nodes (12--1,493 nodes), leaving 10,764 repositories. Very small graphs may reflect trivial projects or incomplete extraction, while very large ones are unpractical even for large context models.

\subsection{Repository Representation}
\label{sec:rep_rep}
Although extremely large repositories are filtered out, the resulting dependency graph, when combined with other contextual documents, can still exceed practical context budgets. To address this, we apply input compression to reduce context length, motivated by prior work showing that longer contexts may degrade model performance even when the additional tokens remain relevant, potentially due to attention dilution and limited effective capacity~\cite{du2025context, modarressi2025nolima}.

Specifically, each repository is encoded in text form using:

\begin{itemize}[leftmargin=1.2em,itemsep=0.15em,topsep=0.2em]
    \item \textbf{Dependency graph:} consists of units, inter-unit dependencies and unit-to-file membership mapping. Dependencies are represented by a unit-level edge list $(\textit{unit}_i, \textit{unit}_j)$ while membership mapping maps each unit to its corresponding files,
$unit \mapsto \{file_1,\allowbreak \dots,\allowbreak file_n\}$;
    \item \textbf{Context documents:} auxiliary files such as external libraries (\eg \texttt{pom.xml}, \texttt{pyproject.toml}), \README files, and others providing additional information (\eg Dockerfile).
\end{itemize}

\begin{figure}[htbp!]
    \centering
    \includegraphics[width=\linewidth]{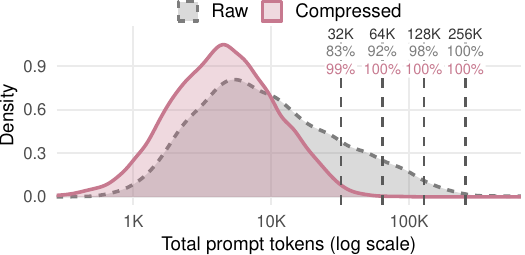}
    \caption{Distribution of prompt tokens per repository for the raw and compressed representations. Dashed lines mark context budgets (32k--256k tokens); the percentages give the share of repositories whose prompt fits within each budget, for the raw (top) and compressed (bottom) representation.}
    \label{fig:token_budget_compare}
\end{figure}

This representation achieves a substantial reduction in context size, up to an order of magnitude for the original, raw graph, while preserving the structural properties most relevant for downstream analysis. Figure~\ref{fig:token_budget_compare} shows the difference in the total token usage between the raw data and the compressed representation for the 10,764 repositories\arxivonly{ (see Appendix~\ref{app:dep_graph_rep} for more details)}.

\section{Proposed Approach}
\label{sec:approach}
\begin{figure}[htbp!]
    \centering
    \includegraphics[width=.8\linewidth]{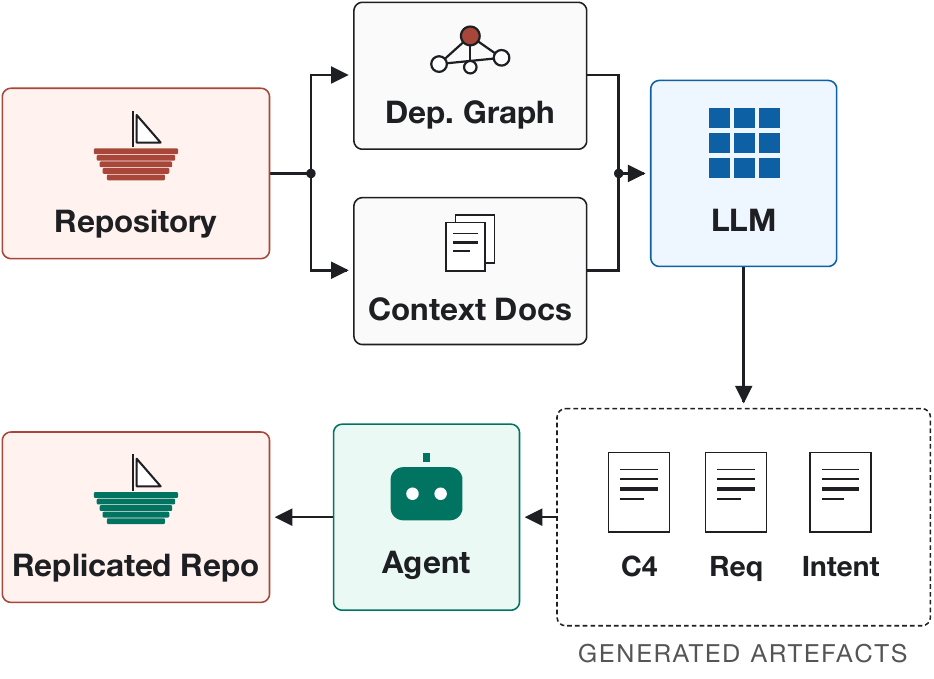}
    \caption{Simplified workflow, a software Ship of Theseus: an LLM recovers design documents from a repository's dependency graph and project files, and an agent rebuilds the library from them.}
    \label{fig:pipeline}
\end{figure}

This section presents the proposed approach for generating aligned early-stage SE artefacts from existing source-code repositories. We then close the loop: a coding agent uses the artefacts to regenerate the repository from scratch, so that the rebuilt code can be compared with the original. Figure~\ref{fig:pipeline} shows a high level view of the approach.

\subsection{Input Representation}

The artefact generation pipelines operate on the repository representation introduced in Section~\ref{sec:rep_rep}. For each project, the input consists of a repository-level dependency graph extracted through static analysis and the available project-level documents. We treat this representation as the fixed context given to the LLM along the specific artefact prompt. 

\subsection{Description of Generated Documents}
This section describes the content and structure of the three artefact types we generate.

\subsubsection{Architecture}
We first generate a software architecture document using the C4 model\footnote{\href{https://c4model.com/}{https://c4model.com/}} as the target format. The C4 model describes a system through progressively more detailed views: System Context, Container, Component, and Code. We exclude the Code level, as it focuses on implementation details rather than architectural structure.

\subsubsection{Requirements}
The next step in our pipeline is to generate software requirements. We generate both functional requirements (FRs), which describe expected system behaviour, and non-functional requirements (NFRs), which capture quality attributes and constraints such as performance, security, reliability, usability, and maintainability.


\subsubsection{Stakeholder Intent}
Stakeholder intent is the initial free-form natural-language specification provided to the code generation system, capturing desired functionalities, constraints, business objectives, user needs, operational expectations, and NFRs such as scalability, security, maintainability, and compliance. It guides the derivation of requirements, architecture, and implementation-related outputs.

\subsection{Artefacts Generation}
We generate each artefact independently with a zero-shot prompt, using the graph and project documents, without conditioning on previously generated intermediate documents.

\subsection{Prompt Design}

We use domain-specific structured inference prompts for all generation steps. Each prompt specifies the target artefact type, output schema, and available input context, but provides no task demonstrations or example outputs.

\subsection{Agent Code Regeneration}
\label{sec:agent_regen}
To test whether the artefacts carry enough of the design to rebuild the system, we close the loop from code to artefacts and back: a coding agent reimplements the repository from scratch, guided by the generated artefacts. The original implementation serves as ground truth: comparing the rebuilt code with it shows how much of the design the artefacts convey.

\section{Validation Methodology}
\label{sec:validation}

Code does not determine a unique architecture, specification or intent, so the artefacts have no single reference. We therefore describe them, rate them with an LLM-as-a-judge approach, and rebuild libraries from them with a coding agent, using the original code as ground truth.

\subsection{Statistics}
\label{sec:val_stat}

We draw 1,000 repositories from the 10,764 of the seed dataset by stratified sampling on language, size and popularity\arxivonly{ (Appendix~\ref{app:eval_subset})}. On these, three reference-free measures describe what each generator produces (Table~\ref{tab:doc_stats}), reported as medians [IQR] over repositories. \emph{Length}, in words, shows how much it writes. \emph{Units named}, the share of code units (modules, files or types; tests excluded) that the requirements or architecture name explicitly, shows how closely the documents refer to the code; it is a lower bound on what they describe. \emph{Diagram validity}, the share of Mermaid C4 diagrams\footnote{\url{https://mermaid.ai/open-source/syntax/c4.html}} that render and of documents whose diagrams all render, checks both the syntax and that relationships connect declared elements.

\subsection{LLM-as-a-Judge}
\label{sec:val_judge}

Three open-weight generators of different families and sizes, Gemma4-31B (\textbf{\mGemma}), Qwen3.5-122B (\textbf{\mQwen}) and GPT-OSS-120B (\textbf{\mGpto}), produce the three artefacts for the same 1,000 repositories, from the same repository representation and prompts. Since having contributors rate reconstructed architecture documentation (as in CIAO~\cite{deluca2026ciao}) is not scalable to more repositories and generators, we use, instead, an LLM evaluator, GPT-6 Luna at its default settings. It judges each artefact in a separate request containing the rubric for its type, the artefact, and the repository representation, which serves as evidence, not as a reference answer. We treat it as a scalable proxy for human assessment, not as ground truth: its ratings are uncalibrated, so we interpret them comparatively.

The evaluator rates each artefact from 1 to 5 on the dimensions of Table~\ref{tab:quality_dimensions}. Each dimension covers one property, so that a defect lowers only one rating, and is adapted to the artefact type; intent, for instance, is judged against what the code supports, since the original stakeholders' intent cannot be recovered. Together, the common dimensions cover the accuracy, completeness, consistency, comprehensibility, redundancy and value questions of the CIAO survey. The evaluator is told to identify defects before scoring: a major defect normally caps a dimension at 3, several minor ones at 4, and a 5 needs positive evidence, not just the absence of errors. 

\begin{table*}[htbp!]
\centering
\small
\caption{Dimensions rated by the evaluator (1--5). Common ones are averaged into the quality score $Q$, specific ones reported separately.}
\label{tab:quality_dimensions}
\begin{tabular}{@{}c l l p{10.3cm} l@{}}
\toprule
 & \textbf{Dimension} & \textbf{Abbr.} & \textbf{Question the evaluator answers} & \textbf{Artefacts} \\
\midrule
\multirow{7}{*}{\rotatebox[origin=c]{90}{\textit{Common}}}
 & Correctness & Corr. & Are the artefact's claims defensible from the repository evidence? & All \\
 & Completeness & Compl. & Does it cover the information that matters at its level of the SDLC? & All \\
 & Abstraction & Abstr. & Is it at the artefact's level of abstraction, not restating the implementation? & All \\
 & Consistency & Cons. & Is it free of internal contradictions? & All \\
 & Clarity & Clar. & Is it clearly written, precise and easy to navigate? & All \\
 & Conciseness & Conc. & Does it avoid repetition and generic filler? & All \\
 & Usefulness & Use. & Could a practitioner rely on it for its intended purpose? & All \\
\midrule
\multirow{3}{*}{\rotatebox[origin=c]{90}{\textit{Specific}}}
 & Relationships & Rel. & Are the interactions and dependencies between architectural elements correct? & Arch. \\
 & Diagram quality & Diag. & Are the diagrams, if any, correct, readable and consistent with the prose? & Arch. \\
 & Section correctness & C4, FR, NFR & Is each C4 view or requirements section present and accurate? & Arch., Req. \\
\bottomrule
\end{tabular}
\end{table*}

We run the evaluation three times and average each rating $r(m,x,a,d)$ (model $m$, repository $x$, artefact type $a$, dimension $d$) over the runs. Unweighted means give the artefact score $Q(m,x,a)$ over the seven common dimensions, the repository score $Q(m,x)$ over the three artefact types, and the model score $Q = Q(m)$ over the repositories; we also report scores per artefact type and dimension. The evaluator's holistic rating of each artefact restates the dimensions, so we exclude it from $Q$ and use its Spearman correlation with $Q(m,x,a)$ as a consistency check.

Reliability across runs was assessed using the intraclass correlation coefficient (ICC), with ICC(3,1) used to evaluate the reliability of a single run and ICC(3,3) used to evaluate the reliability of the mean of three runs. Section~\ref{sec:judge_results} reports the scores (Table~\ref{tab:judge_main}) with bootstrap 95\% confidence intervals, and the reliability. As the artefacts are generated independently, we also check that they form a coherent body: given all three documents, the evaluator links each requirement to the intent items it refines and the architecture elements that realise it. We report the share of intent items and of elements linked to at least one requirement, and the same shares with the requirements of another random repository.

\subsection{Agent-based Repository Reconstruction}
\label{sec:agent_eval}

We run the regeneration on 29 \textsc{Commit0}~\cite{zhao2025commit0} libraries, \textsc{Commit0-Lite} and the others with up to about 8k lines of code, minus two that failed setup; the agent implements each library from scratch and is graded by its own tests. Unlike ProgramBench~\cite{yang2026programbench}, where the agent can run the original executable, \textsc{Commit0} leaves the artefacts and visible tests as the agent's only description of the system, so their effect can be measured. Instead of \textsc{Commit0}'s documentation and stubs, it receives our artefacts, a \emph{contract} naming the modules, classes and functions the tests import, without which they would fail at import, and 30\% of the tests, the \emph{visible} tests, which it can run. The other 70\%, the \emph{held-out} tests, stay hidden from the agent and are used only for grading. The two conditions differ only in the artefacts: \cIntent gives the stakeholder intent alone, i.e., the prompt a developer would write, and \cFull adds the requirements and architecture. These come from two of the three generators, \mGemma and \mQwen, chosen for their contrasting artefacts (Section~\ref{sec:agent_results}). An agent based on Qwen3.8-Flash-Next runs three times per library, condition and generator, under the same budget. Its code is graded in a clean environment, where a test that fails, hangs or does not run counts as failed; we exclude the \AgentRunsExcluded runs whose grading did not complete.

\subsubsection{Metrics}
We compare the rebuilt code on three primary outcomes. For RQ2, \emph{correctness} is the share of held-out tests passed. Tests do not check how the code is organised, so for RQ3 two structural metrics compare the agent's modules with those of the reference implementation. Let $A$, $R$ and $C$ be the modules of the agent's code, of the reference implementation and of the contract. \emph{Module F1} compares the modules not given away by the contract:
\[
P = \frac{|(A \cap R) \setminus C|}{|A \setminus C|}, \enspace
R = \frac{|(A \cap R) \setminus C|}{|R \setminus C|}, \enspace
F_1 = \frac{2\,P\,R}{P + R},
\]
and is undefined when the contract names every module, so we also report it with $C = \emptyset$. \emph{Co-location agreement} checks whether the agent's code and the reference implementations put the same code together, even in differently named modules. It is the adjusted Rand index (ARI) between the module assignments of the public classes and functions both define outside the contract: 1 if identical, about 0 at chance.

Secondary metrics are the agent's input and output tokens, the length of its task prompt, and \emph{artefact modules built} defined as the share of modules named in the requirements and architecture that the agent created.

\subsubsection{Statistical analysis}
We ask whether adding the requirements and architecture (\cFull) improves the rebuilt code over the intent alone (\cIntent). Runs on the same library share its tests and difficulty and are not independent, so the library is the unit of analysis: for each library and condition we average the runs over replicates and generators, giving one paired observation per library. With few pairs, we use the paired Wilcoxon signed-rank test and report the mean difference with a bootstrap 95\% CI and the number of libraries where \cFull is better, tied or worse (Table~\ref{tab:agent_main}). The $p$-values of the three primary outcomes are Holm-adjusted.

Averaging hides the variation between runs and generators. As a robustness check, we also model all runs with \texttt{metric $\sim$ condition $\times$ generator + (1 | library)}, where the random intercept accounts for the repeated runs of each library; the generator effect is estimated from the same model without the interaction. We also repeat the paired test for each generator (Table~\ref{tab:agent_generators}). The interaction $p$-values, \ie whether the benefit of \cFull differs between generators, are secondary.

Both conditions receive the contract and the visible tests, which already reveal part of the structure: the contract names modules the agent must create. \cIntent is therefore a strong structural baseline, and \cFull can only add information about the rest of the code. To see whether this masks its benefit, we relate the per-library gain (\cFull minus \cIntent) to \emph{contract coverage}, $|C|/|R|$, the share of reference modules the contract already names, and to \emph{artefact coverage}, the share of the remaining modules, $R \setminus C$, named in the requirements and architecture. We use Spearman correlations, a median split and a condition $\times$ coverage interaction (Table~\ref{tab:agent_contract}).

Lastly, we ask whether artefacts that name more modules help more. For each of the \CovGainUnits library--generator units we compute the gain averaged over replicates and fit \texttt{gain $\sim$ $\beta \cdot$ predictor + (1 | library)}, where the predictor is artefact coverage, artefact length (tokens of the \cFull task prompt) or both, standardised. $\beta$ is thus the change in gain per standard deviation (SD) of the predictor. With library intercepts absorbing between-library differences, $\beta$ mainly compares the \mGemma and \mQwen artefacts of the same library. 

\section{Results}
\label{sec:results}

\subsection{Artefacts Statistics}
\label{sec:doc_stats}

\begin{table*}[htbp!]
\centering
\small
\caption{Generated documents, medians [IQR]. Units named: share of a repository's code units (modules, files or types; tests excluded) that its requirements or architecture document names. Diagrams: Mermaid C4 diagrams per architecture document, share that render, and share of documents whose diagrams all render.}
\label{tab:doc_stats}
\begin{tabular}{lrrrrrrr}
\toprule
 & \multicolumn{3}{c}{\textbf{Length (words)}} & \textbf{Units} & \multicolumn{3}{c}{\textbf{Mermaid C4 diagrams}} \\
\cmidrule(lr){2-4} \cmidrule(lr){6-8}
\textbf{Model} & \textbf{Intent} & \textbf{Requirements} & \textbf{Architecture} & \textbf{Named (\%)} & \textbf{Per doc.} & \textbf{Valid (\%)} & \textbf{Docs all valid (\%)} \\
\midrule
\mGemma & 468 {\scriptsize [440--494]} & 756 {\scriptsize [655--833]} & 786 {\scriptsize [717--871]} & 13 {\scriptsize [4--33]} & 3 {\scriptsize [3--4]} & 94.3 & 82.4 \\
\mQwen & 701 {\scriptsize [616--800]} & 940 {\scriptsize [790--1,128]} & 1,739 {\scriptsize [1,472--2,069]} & 29 {\scriptsize [10--60]} & 6 {\scriptsize [5--8]} & 83.8 & 53.4 \\
\mGpto & 766 {\scriptsize [685--857]} & 1,481 {\scriptsize [1,338--1,627]} & 1,868 {\scriptsize [1,604--2,174]} & 31 {\scriptsize [13--66]} & 5 {\scriptsize [4--7]} & 57.8 & 20.0 \\
\bottomrule
\end{tabular}
\end{table*}

For each of the 1,000 sampled repositories\arxivonly{ (Appendix~\ref{app:eval_subset})}, every model generates the three artefacts, \DocsTotal documents in total. Table~\ref{tab:doc_stats} reports the measures of Section~\ref{sec:val_stat}.

\subsubsection{Length}
For every model, documents grow longer as the artefact becomes more detailed, from stakeholder intent to requirements to architecture. The models differ mostly in how much they write: \mGemma is the most concise, while \mGpto writes the longest documents, and \mQwen's architecture documents are about twice as long as \mGemma's. Each model's lengths vary little across repositories (narrow IQRs), matching the fixed structure the prompts ask for.

\subsubsection{Units named}
Longer documents name more of the code: \mGemma names a median of \GemmaUnitsNamed\% of the units, against \QwenUnitsNamed\% for \mQwen and \GptoUnitsNamed\% for \mGpto, and almost all names come from the architecture document.

\subsubsection{Diagram validity}
\mGemma produces the fewest diagrams on average per document, but the most reliable ones: \GemmaDiagramsValid\% render, and in \GemmaDocsValid\% of its documents all diagrams render. \mQwen follows with \QwenDiagramsValid\% and \QwenDocsValid\%, and \mGpto trails with \GptoDiagramsValid\% and \GptoDocsValid\%. Generally, most failures (\FailUndefined\% of the \DiagramsFailed failing diagrams) are not syntax errors but relationships to an element that the diagram never declares; in \UndefinedElsewhere\% of these cases the element is declared in another diagram of the same document. Parse errors account for \FailParse\%. 

\begin{takeaway}[label=tk:verbosity]{Generators differ in verbosity, not in form}
All three generators follow the prompted structure, so lengths vary little across repositories; what separates them is how much they write and name. \mGemma writes the shortest documents and names the fewest code units, yet produces the most valid diagrams. Most diagram failures refer to undeclared elements, easily identifiable by render checks.
\end{takeaway}

\subsection{Artefacts Quality}
\label{sec:judge_results}

\begin{table*}[htbp!]
\centering
\small
\setlength{\tabcolsep}{4.2pt}
\caption{Judge scores (1--5) per generator, averaged over runs. $Q$: mean of $Q(m,x)$ over repositories; per artefact: mean of $Q(m,x,a)$; per dimension and artefact-specific: mean rating. Best in bold.}
\label{tab:judge_main}
\begin{tabular}{lcccccccccccccccc}
\toprule
 & & \multicolumn{3}{c}{\textbf{Per artefact}} & \multicolumn{7}{c}{\textbf{Per dimension}} & \multicolumn{5}{c}{\textbf{Artefact-specific}} \\
\cmidrule(lr){3-5} \cmidrule(lr){6-12} \cmidrule(lr){13-17}
\textbf{Model} & \textbf{$Q$} & \textbf{Intent} & \textbf{Req.} & \textbf{Arch.} & \textbf{Corr.} & \textbf{Compl.} & \textbf{Abstr.} & \textbf{Cons.} & \textbf{Clar.} & \textbf{Conc.} & \textbf{Use.} & \textbf{Rel.} & \textbf{Diag.} & \textbf{C4} & \textbf{FR} & \textbf{NFR} \\
\midrule
\mGemma & \textbf{3.48} & \textbf{3.77} & \textbf{3.37} & \textbf{3.30} & \textbf{3.02} & 3.23 & \textbf{2.85} & \textbf{4.31} & \textbf{4.14} & \textbf{3.76} & \textbf{3.02} & \textbf{2.65} & \textbf{2.60} & \textbf{3.03} & \textbf{3.75} & \textbf{2.52} \\
\mQwen & 3.02 & 3.34 & 3.05 & 2.68 & 2.42 & 3.30 & 2.41 & 3.84 & 3.96 & 2.72 & 2.51 & 2.09 & 2.03 & 2.40 & 3.32 & 2.12 \\
\mGpto & 2.84 & 3.10 & 2.76 & 2.67 & 2.11 & \textbf{3.35} & 2.19 & 3.60 & 3.92 & 2.49 & 2.25 & 2.05 & 1.98 & 2.37 & 2.88 & 1.77 \\
\bottomrule
\end{tabular}
\end{table*}

\subsubsection{Quality of the generators}
The LLM-as-a-judge's ratings answer RQ1: the recovered artefacts are clear and internally consistent, but less reliably correct and at the right level of abstraction. \mGemma obtains the highest scores (Table~\ref{tab:judge_main}), with $Q = \GemmaQ$ against \QwenQ for \mQwen and \GptoQ for \mGpto (bootstrap 95\% CIs within $\pm 0.01$), and is rated higher on all three artefact types and on most dimensions. The exception is completeness, where the longer architecture documents of \mQwen and \mGpto score slightly higher. The clearest gap is conciseness, about one point in \mGemma's favour.

\subsubsection{Patterns across generators}
While the absolute ratings are indicative only, their relative pattern is consistent across the three generators (Table~\ref{tab:judge_main}). Correctness, abstraction and usefulness are rated lowest, clarity and consistency highest: the models produce well-formed and internally coherent documents, but they less reliably ground claims in the code and describe the system at the appropriate level of abstraction. Across artefact types, stakeholder intent is rated highest and architecture lowest, and within the requirements, non-functional requirements are rated well below functional ones, suggesting that quality attributes are harder to recover from the dependency graph and context files than behaviour.

Quality depends little on the repository meta-characteristics. Popularity barely affects the overall score (Spearman $|\rho| \le \PopularityRhoAbsMax$) and only weakly intent ($\rho \le \PopularityIntentRhoMax$). Larger repositories yield weaker requirements (mean $\rho$ between repository size and $Q(m,x,a)$: $\SizeRequirementsRhoMean$) and, for two of the three generators, weaker architecture ($\rho$ down to $\SizeArchitectureRhoMin$)\arxivonly{; per stratum in Table~\ref{tab:judge_strata}, Appendix~\ref{app:eval_subset}}.

Lastly, although generated independently, the artefacts are coherent: the requirements refine \AlignIntentCovMin--\AlignIntentCovMax\% of the intent items and are allocated to \AlignArchCovMin--\AlignArchCovMax\% of the architecture elements, against \AlignIntentCovShuffled\% and \AlignArchCovShuffled\% for requirements of another repository.

\subsubsection{Evaluator reliability}
Across three independent runs the evaluator is consistent ($ICC(3,1)=\QIccSingle$, $ICC(3,3)=\QIccMean$), ranks the models identically, and its holistic rating tracks $Q$ (mean Spearman $\rho = \HolisticRhoMean$).

\begin{takeaway}[label=tk:concise_best]{The most concise generator writes the best artefacts}
All generators produce clear, consistent artefacts; the judge consistently rates the most concise best everywhere except completeness, where longer documents gain only slightly. All have room to improve in the same places: grounding claims in the code, staying at the artefact's level of abstraction, and recovering quality attributes, especially in larger repositories.
\end{takeaway}

\subsection{Repository Reconstruction}
\label{sec:agent_results}

\begin{table*}[htbp!]
\centering
\small
\setlength{\tabcolsep}{4.5pt}
\caption{Agent results, averaged over both generators. $\Delta$: \cFull $-$ \cIntent, bold when Wilcoxon $p < .05$, with bootstrap CI. B / T / W: libraries where \cFull is better / tied / worse. Holm: over the three primary outcomes; Mixed: mixed model on all runs.}
\label{tab:agent_main}
\begin{tabular}{lrrrrrr@{\ /\ }c@{\ /\ }lrrr}
\toprule
 & & \multicolumn{4}{c}{\textbf{Mean over libraries}} & \multicolumn{3}{c}{\textbf{Libraries}} & \multicolumn{3}{c}{\textbf{$p$-value}} \\
\cmidrule(lr){3-6} \cmidrule(lr){7-9} \cmidrule(lr){10-12}
\textbf{Metric} & $n$ & \textbf{\cIntent} & \textbf{\cFull} & $\Delta$ & \textbf{95\% CI} & \multicolumn{3}{c}{\textbf{B / T / W}} & \textbf{Wilcoxon} & \textbf{Holm} & \textbf{Mixed} \\
\midrule
\multicolumn{12}{l}{\textit{Correctness}} \\
Held-out tests passed (\%) $\uparrow$ & 29 & 33.7 & 35.8 & \upgood{2.1} & [\ensuremath{-}1.6, \ensuremath{+}5.8] & \textcolor{better}{19} & 1 & \textcolor{worse}{9} & .144 & .160 & .235 \\
\midrule
\multicolumn{12}{l}{\textit{Architecture}} \\
Module F1, non-contract modules $\uparrow$ & 24\rlap{$^{*}$} & 0.54 & 0.63 & \upgood{\textbf{0.09}} & [\ensuremath{+}0.05, \ensuremath{+}0.15] & \textcolor{better}{17} & 0 & \textcolor{worse}{7} & \textbf{.002} & \textbf{.005} & \textbf{$<$.001} \\
\quad precision $\uparrow$ & 24\rlap{$^{*}$} & 0.63 & 0.69 & \upgood{0.06} & [\ensuremath{+}0.01, \ensuremath{+}0.11] & \textcolor{better}{15} & 1 & \textcolor{worse}{8} & .066 & -- & .057 \\
\quad recall $\uparrow$ & 24\rlap{$^{*}$} & 0.51 & 0.63 & \upgood{\textbf{0.12}} & [\ensuremath{+}0.06, \ensuremath{+}0.17] & \textcolor{better}{16} & 3 & \textcolor{worse}{5} & \textbf{.001} & -- & \textbf{$<$.001} \\
Module F1, all modules $\uparrow$ & 29 & 0.77 & 0.81 & \upgood{\textbf{0.04}} & [\ensuremath{+}0.02, \ensuremath{+}0.07] & \textcolor{better}{21} & 0 & \textcolor{worse}{8} & \textbf{.002} & -- & \textbf{.029} \\
Co-location agreement (ARI) $\uparrow$ & 24\rlap{$^{\dagger}$} & 0.92 & 0.88 & \downbad{0.04} & [\ensuremath{-}0.08, \ensuremath{+}0.02] & \textcolor{better}{6} & 5 & \textcolor{worse}{13} & .080 & .160 & .300 \\
\midrule
\multicolumn{12}{l}{\textit{Cost}} \\
Task prompt (k tokens) & 29 & 1.0 & 4.8 & \textbf{\ensuremath{+}3.8} & [\ensuremath{+}3.6, \ensuremath{+}4.0] & 29 & 0 & 0 & \textbf{$<$.001} & -- & \textbf{$<$.001} \\
Agent steps $\downarrow$ & 29 & 869 & 858 & \downgood{11} & [\ensuremath{-}50, \ensuremath{+}30] & \textcolor{better}{13} & 1 & \textcolor{worse}{15} & .678 & -- & .660 \\
Output tokens (k) $\downarrow$ & 29 & 311 & 300 & \downgood{11} & [\ensuremath{-}27, \ensuremath{+}3] & \textcolor{better}{17} & 0 & \textcolor{worse}{12} & .162 & -- & .118 \\
Input tokens (M) $\downarrow$ & 29 & 38.1 & 40.0 & \upbad{\textbf{1.9}} & [\ensuremath{-}0.3, \ensuremath{+}3.8] & \textcolor{better}{9} & 0 & \textcolor{worse}{20} & \textbf{.046} & -- & .068 \\
\midrule
\multicolumn{12}{l}{\textit{Use of the artefacts}} \\
Artefact modules built $\uparrow$ & 22\rlap{$^{\ddagger}$} & 0.58 & 0.70 & \upgood{\textbf{0.12}} & [\ensuremath{+}0.02, \ensuremath{+}0.21] & \textcolor{better}{12} & 4 & \textcolor{worse}{6} & \textbf{.031} & -- & \textbf{.013} \\
\bottomrule
\multicolumn{12}{@{}l}{\footnotesize $^{*}$Without the libraries whose contract names every module.} \\
\multicolumn{12}{@{}l}{\footnotesize $^{\dagger}$Without the libraries whose two implementations share fewer than two classes or functions outside the contract.} \\
\multicolumn{12}{@{}l}{\footnotesize $^{\ddagger}$Without the libraries whose requirements and architecture name no module outside the contract.} \\
\end{tabular}
\end{table*}

The held-out tests answer RQ2 and the structural metrics RQ3: receiving the requirements and architecture does not significantly change how many tests the agent passes, but it makes the agent's module structure significantly closer to the original library's, with the artefacts of both generators. How much of the library the artefacts or the contract name does not change the size of this gain.

The agent rebuilds each library from the artefacts of two contrasting generators (Section~\ref{sec:doc_stats}): \mGemma's are the shortest and name the fewest code units, \mQwen's are about twice as long and name about twice as many. We omit \mGpto, which resembles \mQwen but renders the fewest valid diagrams.

Unless stated otherwise, results \emph{average over the two generators}: a library's value in a condition is its mean over both generators and the three replicates (\AgentRuns valid runs over \AgentLibs libraries). Table~\ref{tab:agent_main} holds every averaged number quoted below, Table~\ref{tab:agent_generators} each generator on its own, and Table~\ref{tab:agent_contract} the moderation analyses, which test whether contract or artefact coverage changes the size of the gain. We quote the Holm-adjusted $p$ of the paired Wilcoxon test for the three primary outcomes and the unadjusted one otherwise; the mixed model on all runs (column Mixed) agrees on all three\arxivonly{. Appendix~\ref{app:agent} gives the secondary metrics (Table~\ref{tab:agent_extra}), each generator's full results (Tables~\ref{tab:agent_all_gemma}--\ref{tab:agent_libraries_qwen}) and per-library results (Table~\ref{tab:agent_libraries})}.

\subsubsection{Correctness}
\cFull does not significantly change how many held-out tests the agent passes: \HeldOutIntent\% in \cIntent and \HeldOutFull\% in \cFull ($\Delta = \HeldOutDiff$ percentage points (pp), 95\% CI [\HeldOutCILow, \HeldOutCIHigh], $p = \HeldOutHolm$). Any effect on correctness is small: the CI excludes losses beyond 2 pp and gains beyond 6 pp. The effect varies strongly between libraries, from \HeldOutMaxLoss to \HeldOutMaxGain pp (Figure~\ref{fig:agent_libraries}(b)), and between runs: the three replicates of the same library and condition have a standard deviation of \HeldOutReplicateSD pp on average. Correctness may be capped by the agent's budget in both conditions alike: \AtCapPct\% of runs use all \AgentMaxSteps steps (\AtCapIntent\% in \cIntent, \AtCapFull\% in \cFull), and the agent passes only about half of the visible tests (\VisiblePassRateIntent\% and \VisiblePassRateFull\%).

\subsubsection{Architecture}

\begin{figure}[t]
    \centering
    \includegraphics[width=\linewidth]{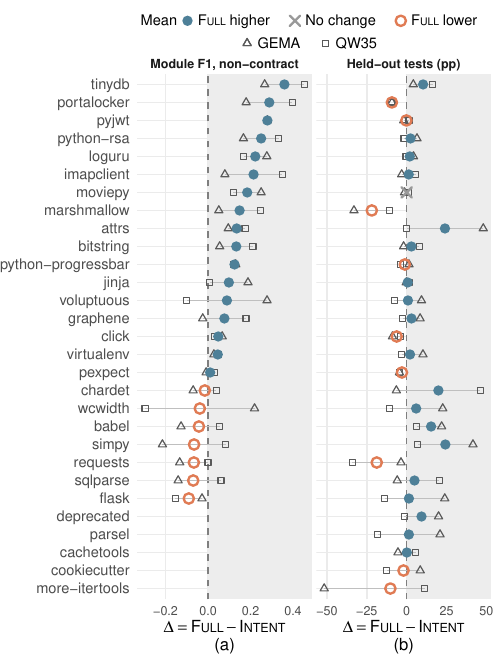}
    \caption{Per-library gain of \cFull over \cIntent in (a) module F1 on the modules the contract does not name and (b) held-out tests passed (pp). Large dots: mean over generators; small markers: each generator. Shaded side: \cFull higher. Sorted by mean module F1 gain; the \ContractFullLibs libraries at the bottom have no module F1, as the contract names every module.}
    \label{fig:agent_libraries}
\end{figure}

\cFull does change how the agent organises the code. The tests alone already reveal much of the structure: in \cIntent, the agent reaches a module F1 of \AllModFIntent over all modules and \ModFIntent over those the contract does not name, which it infers from the intent and the tests. The artefacts add to this. On the modules not given away by the contract, module F1 rises from \ModFIntent to \ModFFull ($\Delta = \ModFDiff$, 95\% CI [\ModFCILow, \ModFCIHigh], $p = \ModFHolm$), closer to the reference on \ModFHigher of \ModFLibs libraries (Figure~\ref{fig:agent_libraries}(a)). The gain comes mostly from recall: the agent creates more of the reference modules, while precision changes less. The agent does use the artefacts: in \cFull it builds \DocRecallFull of the modules they name, against \DocRecallIntent in \cIntent, where many are still implied by the intent or the tests. Counting all modules, including those the contract names, the gain is smaller but significant. Co-location agreement is high in both conditions and does not change significantly ($p = \AriHolm$): when both implementations define the same class or function, they mostly place it in corresponding modules either way. \cFull therefore affects \emph{which} modules exist, not how shared code is grouped.

\subsubsection{Cost}
\cFull gives the agent almost five times as much task text (\TaskTokensFull k against \TaskTokensIntent k tokens) but changes the cost of a run little: output tokens and steps do not change significantly.
Both conditions use most of their step budget (\AgentStepsIntent and \AgentStepsFull steps on average).

\begin{takeaway}[label=tk:structure]{The artefacts shape the structure, not measurably the behaviour}
Adding requirements and architecture barely changes the held-out pass rate or the cost. It changes the decomposition: beyond what the contract and tests reveal, the agent recreates more of the original modules with the artefacts of either generator (recall \ModuleRecallDiff, module F1 \ModFDiff). Recovered design documents thus transfer the part of the design that tests do not specify.
\end{takeaway}

\subsubsection{Generator}
\begin{table}[t]
\centering
\footnotesize
\setlength{\tabcolsep}{3pt}
\caption{Per generator: $\Delta$ (\cFull $-$ \cIntent) over library means, Wilcoxon $p$ in parentheses. Gen.: between generators; Int.: condition $\times$ generator (mixed models, unadjusted).}
\label{tab:agent_generators}
\begin{tabular}{lrrrr}
\toprule
\textbf{Metric} & \textbf{\mGemma} & \textbf{\mQwen} & \textbf{Gen.} & \textbf{Int.} \\
\midrule
Held-out tests (pp) & \ensuremath{+}3.9 (.238) & \ensuremath{-}0.2 (.630) & .832 & .150 \\
Module F1 & \ensuremath{+}0.08 (\textbf{.023}) & \ensuremath{+}0.12 (\textbf{.002}) & .212 & .484 \\
Co-location agreement & \ensuremath{-}0.02 (.468) & \ensuremath{-}0.05 (.159) & .749 & .664 \\
\midrule
Artefact cov. (\%) & 30 & 45 & -- & -- \\
Task prompt, \cFull (k tok.) & 3.3 & 6.3 & -- & -- \\
Artefact modules built, \cFull & 0.70 & 0.69 & -- & -- \\
\bottomrule
\end{tabular}
\end{table}

The gain may depend on the generator, contract coverage, and artefact coverage. However, Table~\ref{tab:agent_generators} and Table~\ref{tab:agent_contract}\arxivonly{ (Figure~\ref{fig:agent_contract} in Appendix~\ref{app:agent} shows contract coverage)} indicate that none significantly affects the gain.

Each primary effect has the same direction with both generators (Table~\ref{tab:agent_generators}): no significant change in correctness or co-location agreement, and a significantly higher module F1. Neither the generator (Gen., \ie whether outcomes differ between generators in both conditions alike) nor its interaction with \cFull (Int., \ie whether the effect of \cFull over \cIntent differs between generators) is significant for any primary outcome. The gain in module F1 is larger with \mQwen (Figure~\ref{fig:agent_libraries}), whose artefacts name more of the library's modules, but the difference is not significant (Int.\ $p = \ModFGenInterP$), and on many libraries the generators disagree on the direction of the effect.

\subsubsection{Contract coverage}
\begin{table}[t]
\centering
\footnotesize
\setlength{\tabcolsep}{3pt}
\caption{Gain of \cFull ($p$ in parentheses). Contract cov.: share of modules the contract names; artefact cov.: share of non-contract modules the artefacts name. $\beta$: within library.}
\label{tab:agent_contract}
\begin{tabular}{llrr}
\toprule
 & & \textbf{Module F1} (24) & \textbf{Held-out (pp)} (29) \\
\midrule
Contract cov. & Spearman $\rho$ & \ensuremath{+}0.29 (.162) & \ensuremath{-}0.24 (.214) \\
 & Gain $<$ / $>$ median & \ensuremath{+}0.05 / \ensuremath{+}0.15 & \ensuremath{+}3.1 / \ensuremath{+}1.2 \\
 & Interaction $p$ & .099 & .230 \\
\midrule
Artefact cov. & Spearman $\rho$ & \ensuremath{+}0.07 (.756) & \ensuremath{-}0.24 (.263) \\
 & $\beta$ per SD & \ensuremath{+}0.011 (.664) & \ensuremath{-}0.5 (.843) \\
 & $\beta$ per SD, length-adj. & \ensuremath{+}0.004 (.884) & \ensuremath{+}0.4 (.882) \\
\midrule
Length & $\beta$ per SD & \ensuremath{+}0.021 (.315) & \ensuremath{-}3.5 (.125) \\
\bottomrule
\end{tabular}
\end{table}

The contracts name between \ContractCovMin\% and \ContractCovMax\% of a library's modules (median \ContractCovMedian\%). If the contract masked the benefit of the artefacts, the gain in module F1 would shrink as contract coverage grows. There is no sign of this: the correlation is positive, though not significant ($\rho = \ModFCovRho$, $p = \ModFCovP$). Splitting the libraries at the median gives the same picture: the gain is \ModFLowCovDiff below and \ModFHighCovDiff above, and the condition $\times$ coverage interaction is not significant ($p = \ModFInteractionP$). For held-out tests the gain is larger below the median (\HeldOutLowCovDiff vs \HeldOutHighCovDiff pp), but neither the correlation ($p = \HeldOutCovP$) nor the interaction ($p = \HeldOutInteractionP$) supports a dependence.

\subsubsection{Artefact coverage and length}
Across libraries, the gain does not grow with the share of non-contract modules that the artefacts name ($\rho = \ModFDocCovRho$, $p = \ModFDocCovP$). Within a library, the generator whose artefacts name more of the modules gains no more module F1: $\beta = \ModFCovGainBeta$ per SD of coverage, i.e.\ per \DocCovSD pp ($p = \ModFCovGainP$), also at equal artefact length ($\beta = \ModFCovGainAdjBeta$, $p = \ModFCovGainAdjP$). Longer artefacts do not help either: $\beta = \ModFLenGainBeta$ per SD of length, i.e.\ per \LengthSD k tokens ($p = \ModFLenGainP$), and \HeldOutLenGainBeta pp of held-out tests passed ($p = \HeldOutLenGainP$).

Within the range of our contracts and generators, what helps the agent is receiving the requirements and architecture, not how many modules they or the contract name.

\begin{takeaway}[label=tk:concise_suffice]{Concise artefacts suffice}
The gain does not require verbose artefacts: those of the concise generator alone bring the module structure significantly closer to the original (Table~\ref{tab:agent_generators}). The verbose generator names about twice as many modules, yet its gain is not significantly larger, and neither the share of modules named nor the length of the artefacts predicts it. Concise artefacts, also cheaper to generate and to read, carry the structure the agent takes up.
\end{takeaway}

\section{Discussion}
\label{sec:discussion}

Given the requirements and architecture, the agent decomposes the system more like the original. As the gain does not grow with the number of modules the artefacts name, it seems to come from the decomposition they describe rather than from the names alone. In frameworks such as MetaGPT~\cite{hong2024metagpt} and ChatDev~\cite{qian2024chatdev}, where a designer agent hands its design to a coder, this suggests that the hand-off steers how the coder decomposes the system, and recovered artefacts can supply such a design when building a new system, and even when original ones don't exist.

The agent passes as many tests in both conditions, but this does not mean that design is irrelevant to behaviour. Tests can only run against an interface the agent is given, such as \textsc{Commit0}'s stubs or our contract, and this interface reveals what will be tested and fixes part of the structure. Whether a design improves correctness when the behaviour is less specified therefore remains open. Few benchmarks target design~\cite{wang2025bench_sdlc}, and test-based ones cannot capture it: in our experiment, the test score alone would have missed the difference. Evaluating design-aware agents requires structural metrics on the part of the code that the interface leaves open. The round trip provides these metrics without human references, using the original code as the reference. It can thus evaluate architecture recovery on any repository with tests, unlike approaches that rely on developers~\cite{deluca2026ciao} or an LLM to judge recovered documentation.

For foundation models, recovered artefacts could add design-level training data from projects without documented designs. Each pair of adjacent abstraction levels, such as requirements and architecture, provides a training task that can support both role-specific agent specialization in multi-agent systems and process-oriented supervision of single agents, encouraging explicit design steps before implementation. For such pairs, what matters is fidelity, i.e., that the artefact describes the code it is paired with, rather than whether its design is good, since the original is only one of many valid designs. Fidelity is what the round trip measures, and even concise artefacts brought the rebuilt structure closer to the original. Fidelity is, however, bounded by what the code reveals: dependency graphs and build files rarely specify performance or security targets. This likely explains why non-functional requirements receive the lowest ratings, consistent with prior extraction work~\cite{miskell2023automated, ouf2025reverse}. Approaches that read the full source~\cite{deluca2026ciao} see more, but their context grows with the lines of code, whereas the dependency graph grows only with the number of units and dependencies.

\section{Threats to Validity}
\label{sec:threats}
We discuss the main threats to the validity of our study and how we mitigate them.

\paragraph{Construct validity}
An LLM judge may not match human judgement and may favour verbose or same-vendor text. We mitigate this with a rubric that anchors the scores to concrete defects, and three runs with high agreement. The ranking also runs against both biases: the judge rates the most concise generator best and the same-vendor \mGpto worst. Moreover, RQ2--RQ3 do not depend on the judge.

\paragraph{Internal validity}
The agent may recall the libraries from pretraining instead of rebuilding them. We mitigate this with a paired design, in which both conditions share the agent, library, contract and test split, so that recall cancels in the comparison, and with a sandbox that gives no access to any copy, as an audit of every run confirms. The agent and \mQwen share a model family, which could favour \mQwen's artefacts; their gain is larger, but not significantly.

\paragraph{External validity}
The round trip uses small Python libraries and a single agent, so its results may not generalise to larger systems, other languages or other agents. We mitigate this by using an established benchmark whose libraries span several domains, and by repeating the experiment with two generators that differ in family and verbosity, which give the same results. The artefacts themselves are not limited to Python: the statistics and the judge cover a sample of 1,000 repositories stratified by language, size and popularity.

\paragraph{Conclusion validity}
With 29 libraries and variable runs, small effects are hard to detect. We mitigate this by running each condition three times, adjusting the primary outcomes for multiple tests, confirming them with a mixed model and reporting confidence intervals; we label the coverage analyses as exploratory.

\section{Conclusions}

LLM agents increasingly design a system before implementing it, either planning their own work or across the roles of a multi-agent framework, yet data connecting intent, requirements and design to code is scarce. We presented a framework that recovers this data from abundant source code, generating aligned stakeholder intent, requirements and architecture at scale, and released the resulting dataset of more than 10k repositories in four languages.

Read as documentation, the recovered artefacts are clear, internally consistent and coherent as a set, with correctness and abstraction leaving the most room for improvement (\textbf{RQ1}). These flaws do not stop them from conveying the design. In a software version of the Ship of Theseus, a coding agent rebuilds each library from its artefacts: compared with \cIntent, \cFull does not significantly change how many held-out tests it passes (\textbf{RQ2}), but brings its module structure significantly closer to the original's (\textbf{RQ3}), carrying design information that the contract and tests do not.

Future work will fine-tune models on the aligned artefacts and code, and test whether the design knowledge they acquire lets them build better-designed systems than their base models, since agents without a design write shallow, monolithic programs~\cite{yang2026programbench}. We also plan to use the artefacts as the hand-offs between the roles of multi-agent frameworks, whether recovered from existing code or produced from scratch by agents trained on the dataset to elicit requirements from the stakeholder intent and to design the architecture. In such pipelines the coder often receives no tests, so we will also rebuild systems without runnable tests, where the artefacts are the only specification and may also improve correctness. We will also test whether the architecture shapes the programs that agents write when the behaviour comes from the original executable and no interface is prescribed, as in ProgramBench~\cite{yang2026programbench}, which isolates the design from the behaviour. Finally, since the round trip needs no human reference, its structural similarity can serve as a reward that trains generators to recover the design a coder needs.

\ifcameraready
\section*{Acknowledgment}
This work has been supported by funding from Swiss State Secretariat for Education, Research and Innovation (SERI), Project number 24.00596 and the HIVEMIND project – Horizon Europe call HORIZON-CL4-2024-DIGITAL-EMERGING-01 under Grant Agreement Number 101189745.
\fi

\bibliographystyle{IEEEtranN}
\bibliography{bibliography}

\ifarxiv
\appendix
\setcounter{table}{0}
\setcounter{figure}{0}
\renewcommand{\thetable}{A\arabic{table}}
\renewcommand{\thefigure}{A\arabic{figure}}
\renewcommand{\theHtable}{A\arabic{table}}
\renewcommand{\theHfigure}{A\arabic{figure}}

\section{Seed Dataset Construction}
\label{app:dataset}

\subsection{List Filtering}
\label{app:list_filter}
The repository collection varies widely in recency (last commit date) and popularity (number of stars), and many repositories are inactive or have limited community engagement. To focus on actively maintained and relevant repositories, we filter repositories using minimum thresholds on star count and the year of the last push.
While the number of stars is not a measure of quality, it serves as a practical filtering heuristic to reduce long-tail noise. Applying a minimum star threshold helps exclude low-visibility repositories with minimal community interest.

Since stricter thresholds reduce dataset size, we analyse coverage across different parameter combinations. The contour plot in Figure~\ref{fig:dataset_coverage} illustrates this trade-off. Based on this analysis, we select $\text{stars} \geq 50$ and $\text{min\_year} = 2021$, yielding a coverage of 46\%, which we consider a good balance.

\begin{figure}[htbp!]
    \centering
    \includegraphics[width=\columnwidth]{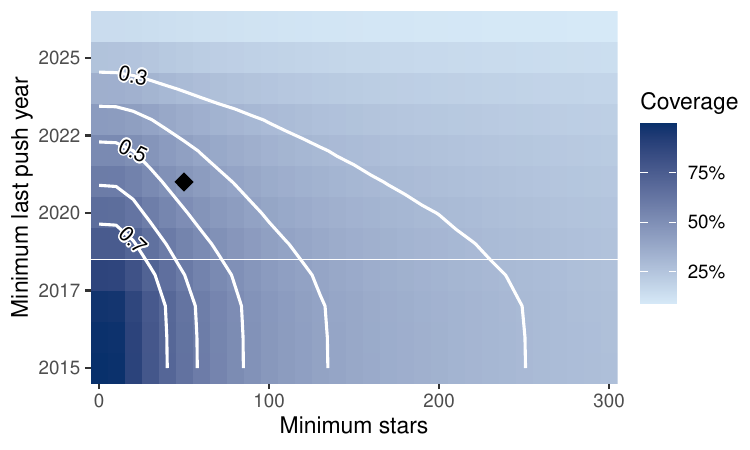}
    \caption{Dataset coverage across different combinations of minimum star counts and minimum last-push year. The contour lines indicate thresholds corresponding to specific percentages of the dataset retained. The black diamond marks the selected configuration: repositories with $\text{stars} \geq 50$ and $\text{min\_year} = 2021$.}
    \label{fig:dataset_coverage}
\end{figure}

\subsection{Graph Filtering}
The analysed projects span a wide range of sizes, from very small repositories (fewer than 5 files) to extremely large ones (10,000 or more files). To ensure meaningful and reliable results, it is necessary to filter out these extremes. Very small projects often indicate incomplete graph extraction, issues in the analysis pipeline, or trivial “toy” repositories that do not reflect realistic structures. Conversely, very large repositories are excluded because they represent atypical, highly complex codebases that fall outside the scope of most practical code generation use cases and can also cause issues with the model's context size. We use percentile-based filtering, with a 10-95th percentile split, resulting in 12 nodes at the lower end and 1493 at the upper end, removing 1,900 projects, and bringing the total to 10,764 repositories.

\subsection{Graph Representation}
\label{app:dep_graph_rep}

The dependency graph extracted by Arcan encodes a wide spectrum of information, ranging from fine-grained file-level metrics (\eg in-degree and out-degree) to higher-level constructs such as class inheritance relationships. However, much of this detail is overly granular and can be systematically compressed without compromising the system’s global structure. In particular, the proposed approach preserves high-level package interactions while abstracting away the underlying file-level dependencies that generate them, thereby reducing redundancy and maintaining a tractable token budget for LLM-based analysis.

Figure~\ref{fig:token_stats_nodes_edges_compression} reports token distribution for the edges and nodes using the \textbf{raw} or the \textbf{compressed} representation. We can see a 5-10 fold reduction achieved by the \textbf{compressed} representation relative to the \textbf{raw}.

\begin{figure}[htbp!]
    \centering
    \includegraphics[width=\linewidth]{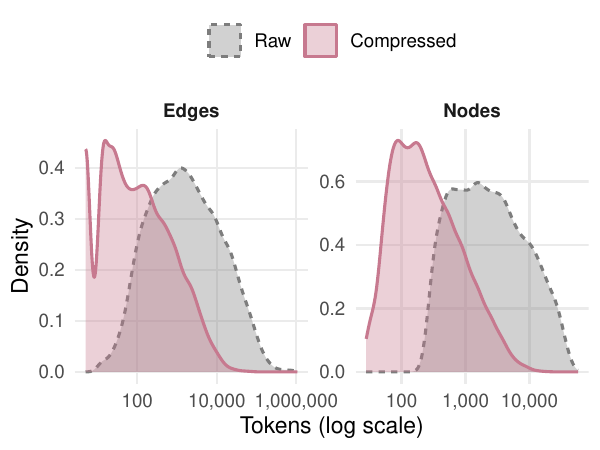}
    \caption{Distribution of the number of tokens for edges and nodes for the raw (dashed outline) and compressed (solid) graph representation.}
    \label{fig:token_stats_nodes_edges_compression}
\end{figure}

\subsection{Additional Context}
To complement the structural information, we extract a set of high-signal, language-native artefacts that capture project functionality, build logic, and runtime behaviour. These artefacts include:
\begin{itemize}
\item dependency files,
\item build files,
\item repository \README files,
\item containerization artefacts (Docker files).
\end{itemize}
Such files are commonly present across software projects and provide structured, machine-readable information about external dependencies, execution flows, runtime settings, and operational environments. Incorporating these artefacts yields a more accurate representation of each repository’s functional characteristics and intended usage.
We avoid using {AGENTS.md} or similar (\eg {CLAUDE.md}) because they contain mostly duplicate information, such as documentation, but not helpful information, such as non-functional requirements~\cite{chatlatanagulchai2025agent}. Therefore, adding duplicate or conflicting information (\eg if the generated {AGENTS.md} contains hallucinations) in our context.


\subsection{Evaluation subset stats}
\label{app:eval_subset}
For the artefact statistics and the LLM-as-a-Judge evaluation, we select a subset of 1,000 repositories from the 10,764 repositories of the filtered set using stratified sampling based on language, binned repository size and binned star count. We use this representative sample to evaluate model performance.

\begin{figure}[htbp!]
    \centering
    \includegraphics[width=\columnwidth]{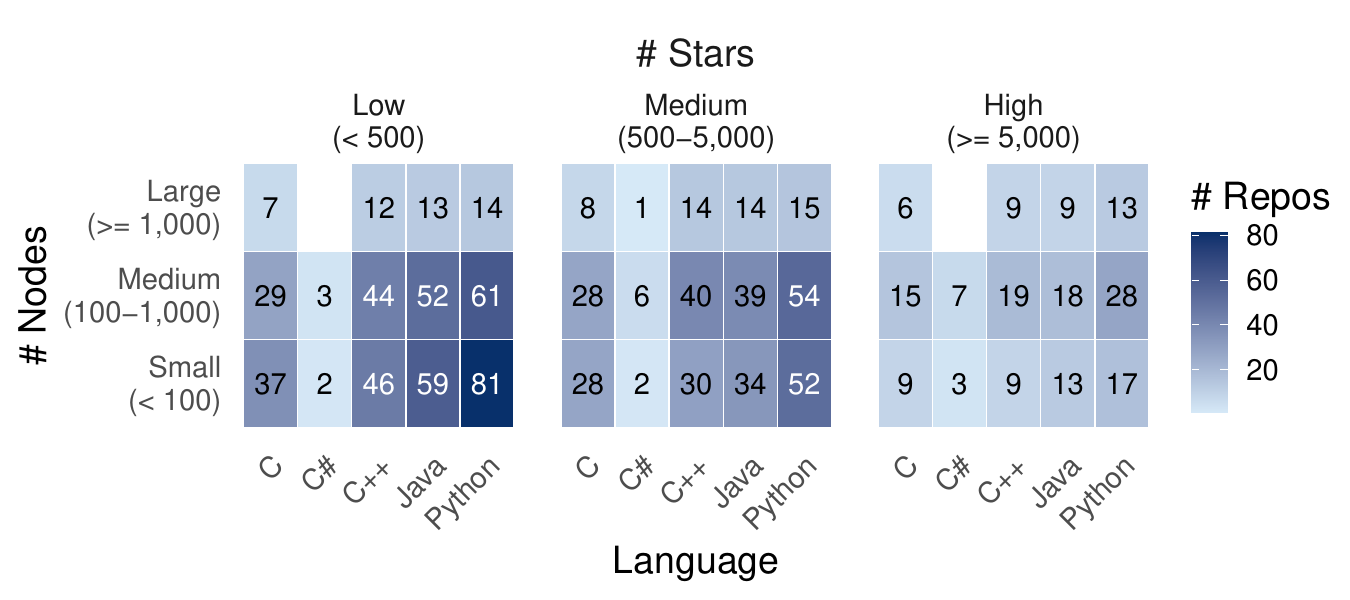}
    \caption{Number of repositories in each stratum}
    \label{fig:placeholder}
\end{figure}

Table~\ref{tab:judge_strata} gives the quality score of each generator per stratum (Section~\ref{sec:judge_results}).

\begin{table*}[htbp!]
\centering
\small
\setlength{\tabcolsep}{2pt}
\caption{Quality score $Q$ per generator and sampling stratum: repository size (nodes of the code graph) and popularity (stars), overall and per artefact type (Intent, Requirements, Architecture). The $\rho$ rows give the Spearman correlation of the score with the number of nodes or stars over all repositories ($^{*}$: $p < 0.05$).}
\label{tab:judge_strata}
\begin{tabular}{@{}llrcccccccccccc@{}}
\toprule
 & &  & \multicolumn{4}{c}{\textbf{\mGemma}} & \multicolumn{4}{c}{\textbf{\mQwen}} & \multicolumn{4}{c}{\textbf{\mGpto}} \\
\cmidrule(lr){4-7} \cmidrule(lr){8-11} \cmidrule(lr){12-15}
\textbf{Stratum} & & \textbf{Repos} & \textbf{$Q$} & \textbf{Intent} & \textbf{Req.} & \textbf{Arch.} & \textbf{$Q$} & \textbf{Intent} & \textbf{Req.} & \textbf{Arch.} & \textbf{$Q$} & \textbf{Intent} & \textbf{Req.} & \textbf{Arch.} \\
\midrule
\multirow{3}{*}{Size} & small ($<$100) & 422 & 3.51 & 3.74 & 3.43 & 3.35 & 3.04 & 3.39 & 3.08 & 2.66 & 2.85 & 3.09 & 2.77 & 2.70 \\
 & medium (100--1k) & 443 & 3.46 & 3.78 & 3.33 & 3.27 & 3.01 & 3.31 & 3.03 & 2.69 & 2.84 & 3.11 & 2.75 & 2.66 \\
 & large ($>$1k) & 135 & 3.42 & 3.80 & 3.26 & 3.21 & 2.99 & 3.29 & 2.98 & 2.68 & 2.82 & 3.09 & 2.74 & 2.62 \\
 & $\rho$ (nodes) &  & $-0.19^{*}$ & $+0.09^{*}$ & $-0.33^{*}$ & $-0.24^{*}$ & $-0.11^{*}$ & $-0.14^{*}$ & $-0.12^{*}$ & $+0.06$ & $-0.08^{*}$ & $+0.03$ & $-0.07^{*}$ & $-0.13^{*}$ \\
\midrule
\multirow{3}{*}{Popularity} & low ($<$500) & 460 & 3.47 & 3.73 & 3.38 & 3.29 & 3.01 & 3.34 & 3.04 & 2.66 & 2.84 & 3.09 & 2.75 & 2.66 \\
 & medium (500--5k) & 365 & 3.48 & 3.78 & 3.36 & 3.31 & 3.03 & 3.32 & 3.05 & 2.71 & 2.85 & 3.09 & 2.76 & 2.68 \\
 & high ($>$5k) & 175 & 3.48 & 3.83 & 3.32 & 3.28 & 3.04 & 3.40 & 3.06 & 2.67 & 2.85 & 3.13 & 2.75 & 2.69 \\
 & $\rho$ (stars) &  & $+0.03$ & $+0.12^{*}$ & $-0.10^{*}$ & $+0.00$ & $+0.06^{*}$ & $+0.06$ & $+0.03$ & $+0.05$ & $+0.07^{*}$ & $+0.07^{*}$ & $+0.01$ & $+0.03$ \\
\bottomrule
\end{tabular}
\end{table*}

\section{Agent Code Generation: Additional Results}
\label{app:agent}

Table~\ref{tab:agent_extra} reports the secondary metrics of the agent evaluation (Section~\ref{sec:agent_eval}), with the same statistics as Table~\ref{tab:agent_main}; Table~\ref{tab:agent_libraries} reports the primary outcomes per library; Figure~\ref{fig:agent_contract} shows the per-library gain against contract coverage. These three average over the two generators. Tables~\ref{tab:agent_all_gemma} and~\ref{tab:agent_all_qwen} give all metrics for each generator on its own, and Tables~\ref{tab:agent_libraries_gemma} and~\ref{tab:agent_libraries_qwen} give the per-library results. The metrics are defined as follows; all are computed per run and averaged over the replicates of a library.

\subsection{Correctness}
\emph{Visible} and \emph{all tests passed}: the share of the visible tests (which the agent could run) and of the whole test suite that pass. The latter is \textsc{Commit0}'s own metric.

\subsection{Architecture}
\emph{Module precision} and \emph{recall}: the two components of module F1, over the modules not named by the contract. \emph{Symbol placement accuracy}: each agent module is mapped to the reference module of the same name or, failing that, to the reference module with which it shares the most classes and functions; the metric is the share of shared classes and functions whose agent module maps to their reference module. \emph{Import-edge precision, recall and F1}: the internal import dependencies between modules (from \texttt{import} statements), with the agent's modules mapped as above and self-loops dropped, compared with the reference's. \emph{Modules created} and \emph{lines of code} (non-blank) measure the size of the agent's implementation; they have no better direction.

\subsection{Cost and behaviour}
\emph{Task text}: the length of the task given to the agent (o200k tokenizer), \ie the size of the treatment. \emph{Agent steps} and \emph{wall time} of the run. \emph{Held-out pp per 100k output tokens}: held-out pass rate (percentage points) divided by the output tokens in units of 100k. \emph{Finished by the agent}: runs that ended because the agent declared the task complete rather than at the step or time limit. \emph{Steps to 50\%/80\% visible passing}: the first step at which one of the agent's own test runs reported at least 50\% / 80\% of the visible tests passing, over the runs that reached it; it measures speed independently of the step limit. \emph{Searches for the reference}: commands looking for the library outside the workspace (none can succeed, as no other copy is readable).

\begin{table*}[htbp!]
\centering
\small
\setlength{\tabcolsep}{4.5pt}
\caption{Agent results, \textbf{\mGemma and \mQwen averaged}: secondary metrics, exploratory ($p$ not corrected). Mean over libraries of each library's mean over generators and replicates; $n$: libraries. \cFull: intent, requirements and C4 architecture; \cIntent: intent only. $\Delta$: \cFull $-$ \cIntent; $\blacktriangle$/$\blacktriangledown$: \cFull higher/lower, a plain signed number when no direction is better; bold when Wilcoxon $p < .05$; 95\% CI: bootstrap over libraries. B / T / W: libraries where \cFull is better / tied / worse (higher / tied / lower when no direction is better). $r$: matched-pairs rank-biserial correlation ($-1$ to 1, positive when \cFull is higher). Wilcoxon: signed-rank test over libraries. Mixed: linear mixed model on the runs, random intercept per library.}
\label{tab:agent_extra}
\begin{tabular}{lrrrrrr@{\ /\ }c@{\ /\ }lrrr}
\toprule
 & & \multicolumn{4}{c}{\textbf{Mean over libraries}} & \multicolumn{3}{c}{\textbf{Libraries}} & \textbf{Effect} & \multicolumn{2}{c}{\textbf{$p$-value}} \\
\cmidrule(lr){3-6} \cmidrule(lr){7-9} \cmidrule(lr){10-10} \cmidrule(lr){11-12}
\textbf{Metric} & $n$ & \textbf{\cIntent} & \textbf{\cFull} & $\Delta$ & \textbf{95\% CI} & \multicolumn{3}{c}{\textbf{B / T / W}} & $r$ & \textbf{Wilcoxon} & \textbf{Mixed} \\
\midrule
\multicolumn{12}{l}{\textit{Correctness}} \\
Visible tests passed (\%) $\uparrow$ & 29 & 49.8 & 51.2 & \upgood{1.4} & [\ensuremath{-}2.8, \ensuremath{+}5.8] & \textcolor{better}{15} & 2 & \textcolor{worse}{12} & \ensuremath{+}0.13 & .578 & .508 \\
All tests passed (\%) $\uparrow$ & 29 & 39.0 & 41.1 & \upgood{2.1} & [\ensuremath{-}1.6, \ensuremath{+}6.0] & \textcolor{better}{17} & 0 & \textcolor{worse}{12} & \ensuremath{+}0.23 & .294 & .252 \\
\midrule
\multicolumn{12}{l}{\textit{Architecture}} \\
Symbol placement accuracy $\uparrow$ & 27 & 0.92 & 0.92 & 0.00 & [\ensuremath{-}0.04, \ensuremath{+}0.04] & \textcolor{better}{10} & 3 & \textcolor{worse}{14} & \ensuremath{-}0.05 & .833 & .838 \\
Import-edge precision $\uparrow$ & 29 & 0.70 & 0.70 & 0.00 & [\ensuremath{-}0.04, \ensuremath{+}0.04] & \textcolor{better}{12} & 0 & \textcolor{worse}{17} & \ensuremath{-}0.24 & .275 & .769 \\
Import-edge recall $\uparrow$ & 29 & 0.57 & 0.63 & \upgood{\textbf{0.06}} & [\ensuremath{+}0.02, \ensuremath{+}0.09] & \textcolor{better}{21} & 1 & \textcolor{worse}{7} & \ensuremath{+}0.60 & \textbf{.005} & \textbf{.008} \\
Import-edge F1 $\uparrow$ & 29 & 0.60 & 0.64 & \upgood{0.04} & [0.00, \ensuremath{+}0.07] & \textcolor{better}{18} & 0 & \textcolor{worse}{11} & \ensuremath{+}0.38 & .076 & .081 \\
Modules created & 29 & 18.4 & 19.3 & \ensuremath{+}0.9 & [0.0, \ensuremath{+}2.0] & 17 & 1 & 11 & \ensuremath{+}0.33 & .121 & .074 \\
Lines of code (k) & 29 & 4.62 & 4.49 & \ensuremath{-}0.13 & [\ensuremath{-}0.47, \ensuremath{+}0.18] & 14 & 0 & 15 & \ensuremath{-}0.06 & .798 & .639 \\
\midrule
\multicolumn{12}{l}{\textit{Cost and behaviour}} \\
Wall time (h) $\downarrow$ & 29 & 2.43 & 2.32 & \downgood{0.11} & [\ensuremath{-}0.27, \ensuremath{+}0.03] & \textcolor{better}{16} & 0 & \textcolor{worse}{13} & \ensuremath{-}0.27 & .213 & .145 \\
Held-out pp per 100k output tokens $\uparrow$ & 29 & 12.78 & 16.10 & \upgood{3.32} & [\ensuremath{+}0.55, \ensuremath{+}7.08] & \textcolor{better}{19} & 0 & \textcolor{worse}{10} & \ensuremath{+}0.42 & .051 & \textbf{.012} \\
Finished by the agent (\%) $\uparrow$ & 29 & 15.2 & 20.1 & \upgood{4.9} & [\ensuremath{-}0.9, \ensuremath{+}11.2] & \textcolor{better}{9} & 14 & \textcolor{worse}{6} & \ensuremath{+}0.36 & .214 & .133 \\
Steps to 50\% visible passing $\downarrow$ & 18 & 314 & 310 & \downgood{4} & [\ensuremath{-}53, \ensuremath{+}43] & \textcolor{better}{9} & 0 & \textcolor{worse}{9} & \ensuremath{-}0.02 & .966 & .705 \\
Steps to 80\% visible passing $\downarrow$ & 14 & 398 & 414 & \upbad{16} & [\ensuremath{-}57, \ensuremath{+}91] & \textcolor{better}{7} & 0 & \textcolor{worse}{7} & \ensuremath{+}0.10 & .761 & .452 \\
Searches for the reference $\downarrow$ & 29 & 5.6 & 4.8 & \downgood{0.8} & [\ensuremath{-}1.6, 0.0] & \textcolor{better}{17} & 1 & \textcolor{worse}{11} & \ensuremath{-}0.36 & .094 & .054 \\
\bottomrule
\end{tabular}
\end{table*}

\begin{table*}[htbp!]
\centering
\small
\caption{Per-library results, \textbf{\mGemma and \mQwen averaged}: mean over generators and replicates. \cFull: intent, requirements and C4 architecture; \cIntent: intent only. $^{\dagger}$\textsc{Commit0-Lite}. --: undefined, because the contract names every module of the library.}
\label{tab:agent_libraries}
\begin{tabular}{lrrrrrrrrrr}
\toprule
 & & \multicolumn{3}{c}{\textbf{Held-out passed (\%)}} & \multicolumn{3}{c}{\textbf{Module F1, non-contract}} & \multicolumn{3}{c}{\textbf{Module F1, all}} \\
\cmidrule(lr){3-5} \cmidrule(lr){6-8} \cmidrule(lr){9-11}
\textbf{Library} & \textbf{Held-out tests} & \textbf{\cIntent} & \textbf{\cFull} & $\Delta$ & \textbf{\cIntent} & \textbf{\cFull} & $\Delta$ & \textbf{\cIntent} & \textbf{\cFull} & $\Delta$ \\
\midrule
attrs & 960 & 0.0 & 24.0 & \upgood{24.0} & 0.77 & 0.90 & \upgood{0.13} & 0.79 & 0.86 & \upgood{0.07} \\
babel$^{\dagger}$ & 4387 & 5.7 & 21.1 & \upgood{15.4} & 0.17 & 0.12 & \downbad{0.05} & 0.41 & 0.51 & \upgood{0.10} \\
bitstring & 590 & 8.4 & 11.4 & \upgood{3.0} & 0.21 & 0.34 & \upgood{0.13} & 0.63 & 0.68 & \upgood{0.05} \\
cachetools$^{\dagger}$ & 159 & 51.0 & 51.2 & \upgood{0.2} & -- & -- & -- & 0.59 & 0.74 & \upgood{0.15} \\
chardet$^{\dagger}$ & 376 & 16.4 & 36.3 & \upgood{19.9} & 0.44 & 0.43 & \downbad{0.01} & 0.49 & 0.47 & \downbad{0.02} \\
click & 420 & 29.0 & 22.7 & \downbad{6.3} & 0.81 & 0.86 & \upgood{0.05} & 0.95 & 0.96 & \upgood{0.01} \\
cookiecutter$^{\dagger}$ & 243 & 35.1 & 33.1 & \downbad{2.0} & -- & -- & -- & 0.94 & 0.95 & \upgood{0.01} \\
deprecated$^{\dagger}$ & 120 & 36.7 & 46.0 & \upgood{9.3} & -- & -- & -- & 0.86 & 0.93 & \upgood{0.07} \\
flask & 339 & 23.7 & 25.1 & \upgood{1.4} & 0.55 & 0.46 & \downbad{0.09} & 0.83 & 0.69 & \downbad{0.14} \\
graphene & 293 & 7.5 & 10.5 & \upgood{3.0} & 0.72 & 0.79 & \upgood{0.07} & 0.77 & 0.84 & \upgood{0.07} \\
imapclient$^{\dagger}$ & 192 & 36.4 & 37.7 & \upgood{1.3} & 0.29 & 0.51 & \upgood{0.22} & 0.72 & 0.77 & \upgood{0.05} \\
jinja$^{\dagger}$ & 604 & 0.5 & 1.2 & \upgood{0.7} & 0.37 & 0.46 & \upgood{0.09} & 0.43 & 0.46 & \upgood{0.03} \\
loguru & 1046 & 4.4 & 6.3 & \upgood{1.9} & 0.30 & 0.52 & \upgood{0.22} & 0.47 & 0.67 & \upgood{0.20} \\
marshmallow$^{\dagger}$ & 858 & 74.8 & 52.9 & \downbad{21.9} & 0.73 & 0.88 & \upgood{0.15} & 0.90 & 0.96 & \upgood{0.06} \\
more-itertools & 464 & 54.7 & 44.5 & \downbad{10.2} & -- & -- & -- & 0.75 & 0.92 & \upgood{0.17} \\
moviepy & 74 & 48.4 & 48.4 & 0.0 & 0.45 & 0.63 & \upgood{0.18} & 0.67 & 0.84 & \upgood{0.17} \\
parsel$^{\dagger}$ & 145 & 28.7 & 30.1 & \upgood{1.4} & -- & -- & -- & 0.95 & 0.98 & \upgood{0.03} \\
pexpect & 180 & 35.4 & 32.5 & \downbad{2.9} & 0.64 & 0.65 & \upgood{0.01} & 0.85 & 0.80 & \downbad{0.05} \\
portalocker$^{\dagger}$ & 18 & 37.0 & 27.8 & \downbad{9.2} & 0.44 & 0.73 & \upgood{0.29} & 0.83 & 0.91 & \upgood{0.08} \\
pyjwt$^{\dagger}$ & 182 & 75.0 & 74.8 & \downbad{0.2} & 0.58 & 0.86 & \upgood{0.28} & 0.89 & 0.96 & \upgood{0.07} \\
python-progressbar & 257 & 19.0 & 17.8 & \downbad{1.2} & 0.19 & 0.31 & \upgood{0.12} & 0.63 & 0.68 & \upgood{0.05} \\
python-rsa & 62 & 66.9 & 69.4 & \upgood{2.5} & 0.33 & 0.58 & \upgood{0.25} & 0.97 & 0.97 & 0.00 \\
requests & 413 & 88.2 & 69.4 & \downbad{18.8} & 0.92 & 0.85 & \downbad{0.07} & 0.98 & 0.96 & \downbad{0.02} \\
simpy$^{\dagger}$ & 101 & 19.1 & 43.4 & \upgood{24.3} & 0.83 & 0.77 & \downbad{0.06} & 0.92 & 0.86 & \downbad{0.06} \\
sqlparse & 287 & 17.6 & 22.5 & \upgood{4.9} & 0.68 & 0.61 & \downbad{0.07} & 0.81 & 0.78 & \downbad{0.03} \\
tinydb$^{\dagger}$ & 144 & 48.7 & 59.0 & \upgood{10.3} & 0.38 & 0.74 & \upgood{0.36} & 0.85 & 0.94 & \upgood{0.09} \\
virtualenv & 217 & 22.0 & 24.1 & \upgood{2.1} & 0.48 & 0.52 & \upgood{0.04} & 0.73 & 0.78 & \upgood{0.05} \\
voluptuous$^{\dagger}$ & 104 & 34.5 & 35.3 & \upgood{0.8} & 0.74 & 0.83 & \upgood{0.09} & 0.90 & 0.94 & \upgood{0.04} \\
wcwidth$^{\dagger}$ & 28 & 53.6 & 59.5 & \upgood{5.9} & 0.81 & 0.78 & \downbad{0.03} & 0.76 & 0.75 & \downbad{0.01} \\
\midrule
Mean &  & 33.7 & 35.8 & \upgood{2.1} & 0.54 & 0.63 & \upgood{\textbf{0.09}} & 0.77 & 0.81 & \upgood{\textbf{0.04}} \\
\bottomrule
\end{tabular}
\end{table*}

\begin{figure}[htbp!]
    \centering
    \includegraphics[width=\linewidth]{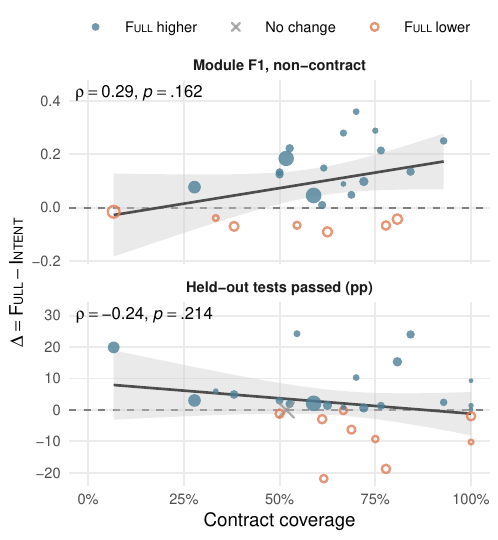}
    \caption{Averaged over both generators: per-library difference between \cFull and \cIntent against contract coverage, the share of the library's modules that the contract names. One point per library, sized by its number of modules (filled: \cFull higher; hollow: \cFull lower; cross: no change); line: least-squares fit with 95\% band. Module F1 is undefined for the \ContractFullLibs libraries whose contract names every module.}
    \label{fig:agent_contract}
\end{figure}

\begin{table*}[htbp!]
\centering
\small
\setlength{\tabcolsep}{4.5pt}
\caption{Agent results, \textbf{\mGemma only}: all metrics, exploratory ($p$ not corrected). Mean over libraries of each library's mean over replicates; $n$: libraries. \cFull: intent, requirements and C4 architecture; \cIntent: intent only. $\Delta$: \cFull $-$ \cIntent; $\blacktriangle$/$\blacktriangledown$: \cFull higher/lower, a plain signed number when no direction is better; bold when Wilcoxon $p < .05$; 95\% CI: bootstrap over libraries. B / T / W: libraries where \cFull is better / tied / worse (higher / tied / lower when no direction is better). $r$: matched-pairs rank-biserial correlation ($-1$ to 1, positive when \cFull is higher). Wilcoxon: signed-rank test over libraries. Mixed: linear mixed model on the runs, random intercept per library.}
\label{tab:agent_all_gemma}
\begin{tabular}{lrrrrrr@{\ /\ }c@{\ /\ }lrrr}
\toprule
 & & \multicolumn{4}{c}{\textbf{Mean over libraries}} & \multicolumn{3}{c}{\textbf{Libraries}} & \textbf{Effect} & \multicolumn{2}{c}{\textbf{$p$-value}} \\
\cmidrule(lr){3-6} \cmidrule(lr){7-9} \cmidrule(lr){10-10} \cmidrule(lr){11-12}
\textbf{Metric} & $n$ & \textbf{\cIntent} & \textbf{\cFull} & $\Delta$ & \textbf{95\% CI} & \multicolumn{3}{c}{\textbf{B / T / W}} & $r$ & \textbf{Wilcoxon} & \textbf{Mixed} \\
\midrule
\multicolumn{12}{l}{\textit{Correctness}} \\
Held-out tests passed (\%) $\uparrow$ & 29 & 32.6 & 36.5 & \upgood{3.9} & [\ensuremath{-}3.2, \ensuremath{+}10.9] & \textcolor{better}{15} & 0 & \textcolor{worse}{14} & \ensuremath{+}0.26 & .238 & .061 \\
\midrule
\multicolumn{12}{l}{\textit{Architecture}} \\
Module F1, non-contract modules $\uparrow$ & 24\rlap{$^{*}$} & 0.56 & 0.64 & \upgood{\textbf{0.08}} & [\ensuremath{+}0.02, \ensuremath{+}0.14] & \textcolor{better}{16} & 0 & \textcolor{worse}{8} & \ensuremath{+}0.53 & \textbf{.023} & \textbf{.035} \\
\quad precision $\uparrow$ & 24\rlap{$^{*}$} & 0.67 & 0.71 & \upgood{0.04} & [\ensuremath{-}0.04, \ensuremath{+}0.11] & \textcolor{better}{14} & 2 & \textcolor{worse}{8} & \ensuremath{+}0.30 & .217 & .434 \\
\quad recall $\uparrow$ & 24\rlap{$^{*}$} & 0.54 & 0.63 & \upgood{\textbf{0.09}} & [\ensuremath{+}0.03, \ensuremath{+}0.16] & \textcolor{better}{14} & 2 & \textcolor{worse}{8} & \ensuremath{+}0.58 & \textbf{.017} & \textbf{.009} \\
Module F1, all modules $\uparrow$ & 29 & 0.78 & 0.82 & \upgood{0.04} & [\ensuremath{+}0.01, \ensuremath{+}0.08] & \textcolor{better}{17} & 0 & \textcolor{worse}{12} & \ensuremath{+}0.37 & .088 & .107 \\
Co-location agreement (ARI) $\uparrow$ & 23\rlap{$^{\dagger}$} & 0.90 & 0.88 & \downbad{0.02} & [\ensuremath{-}0.08, \ensuremath{+}0.04] & \textcolor{better}{6} & 5 & \textcolor{worse}{12} & \ensuremath{-}0.20 & .468 & .700 \\
\midrule
\multicolumn{12}{l}{\textit{Cost}} \\
Task prompt (k tokens) & 29 & 0.8 & 3.3 & \textbf{\ensuremath{+}2.5} & [\ensuremath{+}2.4, \ensuremath{+}2.6] & 29 & 0 & 0 & \ensuremath{+}1.00 & \textbf{$<$.001} & \textbf{$<$.001} \\
Agent steps $\downarrow$ & 29 & 851 & 853 & \upbad{2} & [\ensuremath{-}60, \ensuremath{+}62] & \textcolor{better}{11} & 6 & \textcolor{worse}{12} & \ensuremath{+}0.05 & .820 & .802 \\
Output tokens (k) $\downarrow$ & 29 & 312 & 302 & \downgood{10} & [\ensuremath{-}38, \ensuremath{+}15] & \textcolor{better}{14} & 0 & \textcolor{worse}{15} & \ensuremath{-}0.11 & .609 & .482 \\
Input tokens (M) $\downarrow$ & 29 & 38.2 & 38.9 & \upbad{0.7} & [\ensuremath{-}2.3, \ensuremath{+}3.4] & \textcolor{better}{13} & 0 & \textcolor{worse}{16} & \ensuremath{+}0.18 & .417 & .408 \\
\midrule
\multicolumn{12}{l}{\textit{Use of the artefacts}} \\
Artefact modules built $\uparrow$ & 18\rlap{$^{\ddagger}$} & 0.65 & 0.70 & \upgood{0.05} & [\ensuremath{-}0.08, \ensuremath{+}0.17] & \textcolor{better}{7} & 5 & \textcolor{worse}{6} & \ensuremath{+}0.27 & .413 & .323 \\
\midrule
\multicolumn{12}{l}{\textit{Correctness}} \\
Visible tests passed (\%) $\uparrow$ & 29 & 48.8 & 52.0 & \upgood{3.2} & [\ensuremath{-}5.8, \ensuremath{+}12.0] & \textcolor{better}{15} & 1 & \textcolor{worse}{13} & \ensuremath{+}0.18 & .412 & .234 \\
All tests passed (\%) $\uparrow$ & 29 & 37.9 & 41.6 & \upgood{3.7} & [\ensuremath{-}4.1, \ensuremath{+}11.0] & \textcolor{better}{15} & 0 & \textcolor{worse}{14} & \ensuremath{+}0.21 & .325 & .107 \\
\midrule
\multicolumn{12}{l}{\textit{Architecture}} \\
Symbol placement accuracy $\uparrow$ & 26 & 0.90 & 0.90 & 0.00 & [\ensuremath{-}0.10, \ensuremath{+}0.08] & \textcolor{better}{13} & 4 & \textcolor{worse}{9} & \ensuremath{+}0.17 & .503 & .418 \\
Import-edge precision $\uparrow$ & 29 & 0.71 & 0.70 & \downbad{0.01} & [\ensuremath{-}0.05, \ensuremath{+}0.04] & \textcolor{better}{11} & 1 & \textcolor{worse}{17} & \ensuremath{-}0.20 & .362 & .788 \\
Import-edge recall $\uparrow$ & 29 & 0.57 & 0.65 & \upgood{\textbf{0.08}} & [\ensuremath{+}0.02, \ensuremath{+}0.14] & \textcolor{better}{18} & 1 & \textcolor{worse}{10} & \ensuremath{+}0.47 & \textbf{.030} & \textbf{.007} \\
Import-edge F1 $\uparrow$ & 29 & 0.60 & 0.65 & \upgood{0.05} & [0.00, \ensuremath{+}0.10] & \textcolor{better}{17} & 1 & \textcolor{worse}{11} & \ensuremath{+}0.31 & .157 & .056 \\
Modules created & 29 & 18.4 & 18.9 & \ensuremath{+}0.5 & [\ensuremath{-}0.7, \ensuremath{+}1.5] & 20 & 0 & 9 & \ensuremath{+}0.31 & .138 & .383 \\
Lines of code (k) & 29 & 4.67 & 4.39 & \ensuremath{-}0.28 & [\ensuremath{-}0.98, \ensuremath{+}0.26] & 14 & 0 & 15 & \ensuremath{-}0.13 & .550 & .623 \\
\midrule
\multicolumn{12}{l}{\textit{Cost and behaviour}} \\
Wall time (h) $\downarrow$ & 29 & 2.49 & 2.28 & \downgood{0.21} & [\ensuremath{-}0.45, \ensuremath{+}0.02] & \textcolor{better}{18} & 0 & \textcolor{worse}{11} & \ensuremath{-}0.30 & .162 & .107 \\
Held-out pp per 100k output tokens $\uparrow$ & 29 & 12.09 & 17.38 & \upgood{5.29} & [\ensuremath{+}0.65, \ensuremath{+}11.97] & \textcolor{better}{14} & 0 & \textcolor{worse}{15} & \ensuremath{+}0.31 & .143 & \textbf{.007} \\
Finished by the agent (\%) $\uparrow$ & 28 & 14.3 & 20.2 & \upgood{5.9} & [\ensuremath{-}3.6, \ensuremath{+}16.7] & \textcolor{better}{6} & 18 & \textcolor{worse}{4} & \ensuremath{+}0.27 & .498 & .440 \\
Steps to 50\% visible passing $\downarrow$ & 17 & 319 & 282 & \downgood{37} & [\ensuremath{-}113, \ensuremath{+}42] & \textcolor{better}{11} & 0 & \textcolor{worse}{6} & \ensuremath{-}0.29 & .306 & .176 \\
Steps to 80\% visible passing $\downarrow$ & 12 & 373 & 386 & \upbad{13} & [\ensuremath{-}62, \ensuremath{+}87] & \textcolor{better}{5} & 0 & \textcolor{worse}{7} & \ensuremath{+}0.08 & .850 & .975 \\
Searches for the reference $\downarrow$ & 29 & 5.9 & 4.7 & \downgood{1.2} & [\ensuremath{-}2.5, \ensuremath{-}0.1] & \textcolor{better}{18} & 0 & \textcolor{worse}{11} & \ensuremath{-}0.34 & .114 & .066 \\
\bottomrule
\multicolumn{12}{@{}l}{\footnotesize $^{*}$Without the libraries whose contract names every module.} \\
\multicolumn{12}{@{}l}{\footnotesize $^{\dagger}$Without the libraries whose two implementations share fewer than two classes or functions outside the contract.} \\
\multicolumn{12}{@{}l}{\footnotesize $^{\ddagger}$Without the libraries whose requirements and architecture name no module outside the contract.} \\
\end{tabular}
\end{table*}

\begin{table*}[htbp!]
\centering
\small
\setlength{\tabcolsep}{4.5pt}
\caption{Agent results, \textbf{\mQwen only}: all metrics, exploratory ($p$ not corrected). Mean over libraries of each library's mean over replicates; $n$: libraries. \cFull: intent, requirements and C4 architecture; \cIntent: intent only. $\Delta$: \cFull $-$ \cIntent; $\blacktriangle$/$\blacktriangledown$: \cFull higher/lower, a plain signed number when no direction is better; bold when Wilcoxon $p < .05$; 95\% CI: bootstrap over libraries. B / T / W: libraries where \cFull is better / tied / worse (higher / tied / lower when no direction is better). $r$: matched-pairs rank-biserial correlation ($-1$ to 1, positive when \cFull is higher). Wilcoxon: signed-rank test over libraries. Mixed: linear mixed model on the runs, random intercept per library.}
\label{tab:agent_all_qwen}
\begin{tabular}{lrrrrrr@{\ /\ }c@{\ /\ }lrrr}
\toprule
 & & \multicolumn{4}{c}{\textbf{Mean over libraries}} & \multicolumn{3}{c}{\textbf{Libraries}} & \textbf{Effect} & \multicolumn{2}{c}{\textbf{$p$-value}} \\
\cmidrule(lr){3-6} \cmidrule(lr){7-9} \cmidrule(lr){10-10} \cmidrule(lr){11-12}
\textbf{Metric} & $n$ & \textbf{\cIntent} & \textbf{\cFull} & $\Delta$ & \textbf{95\% CI} & \multicolumn{3}{c}{\textbf{B / T / W}} & $r$ & \textbf{Wilcoxon} & \textbf{Mixed} \\
\midrule
\multicolumn{12}{l}{\textit{Correctness}} \\
Held-out tests passed (\%) $\uparrow$ & 29 & 35.1 & 34.9 & \downbad{0.2} & [\ensuremath{-}5.0, \ensuremath{+}5.1] & \textcolor{better}{12} & 1 & \textcolor{worse}{16} & \ensuremath{-}0.11 & .630 & .876 \\
\midrule
\multicolumn{12}{l}{\textit{Architecture}} \\
Module F1, non-contract modules $\uparrow$ & 24\rlap{$^{*}$} & 0.51 & 0.63 & \upgood{\textbf{0.12}} & [\ensuremath{+}0.05, \ensuremath{+}0.19] & \textcolor{better}{20} & 1 & \textcolor{worse}{3} & \ensuremath{+}0.71 & \textbf{.002} & \textbf{.002} \\
\quad precision $\uparrow$ & 24\rlap{$^{*}$} & 0.59 & 0.67 & \upgood{0.08} & [\ensuremath{+}0.01, \ensuremath{+}0.17] & \textcolor{better}{14} & 2 & \textcolor{worse}{8} & \ensuremath{+}0.42 & .082 & .055 \\
\quad recall $\uparrow$ & 24\rlap{$^{*}$} & 0.49 & 0.62 & \upgood{\textbf{0.13}} & [\ensuremath{+}0.06, \ensuremath{+}0.21] & \textcolor{better}{18} & 2 & \textcolor{worse}{4} & \ensuremath{+}0.70 & \textbf{.004} & \textbf{$<$.001} \\
Module F1, all modules $\uparrow$ & 29 & 0.76 & 0.80 & \upgood{\textbf{0.04}} & [0.00, \ensuremath{+}0.08] & \textcolor{better}{22} & 1 & \textcolor{worse}{6} & \ensuremath{+}0.56 & \textbf{.008} & .149 \\
Co-location agreement (ARI) $\uparrow$ & 22\rlap{$^{\dagger}$} & 0.92 & 0.87 & \downbad{0.05} & [\ensuremath{-}0.12, \ensuremath{+}0.03] & \textcolor{better}{6} & 5 & \textcolor{worse}{11} & \ensuremath{-}0.40 & .159 & .230 \\
\midrule
\multicolumn{12}{l}{\textit{Cost}} \\
Task prompt (k tokens) & 29 & 1.1 & 6.3 & \textbf{\ensuremath{+}5.2} & [\ensuremath{+}4.8, \ensuremath{+}5.5] & 29 & 0 & 0 & \ensuremath{+}1.00 & \textbf{$<$.001} & \textbf{$<$.001} \\
Agent steps $\downarrow$ & 29 & 888 & 863 & \downgood{25} & [\ensuremath{-}91, \ensuremath{+}38] & \textcolor{better}{13} & 6 & \textcolor{worse}{10} & \ensuremath{-}0.17 & .501 & .447 \\
Output tokens (k) $\downarrow$ & 29 & 312 & 296 & \downgood{16} & [\ensuremath{-}36, \ensuremath{+}5] & \textcolor{better}{16} & 0 & \textcolor{worse}{13} & \ensuremath{-}0.31 & .156 & .124 \\
Input tokens (M) $\downarrow$ & 29 & 38.4 & 41.0 & \upbad{2.6} & [\ensuremath{-}0.5, \ensuremath{+}5.8] & \textcolor{better}{12} & 0 & \textcolor{worse}{17} & \ensuremath{+}0.29 & .183 & .088 \\
\midrule
\multicolumn{12}{l}{\textit{Use of the artefacts}} \\
Artefact modules built $\uparrow$ & 21\rlap{$^{\ddagger}$} & 0.55 & 0.69 & \upgood{\textbf{0.14}} & [\ensuremath{+}0.05, \ensuremath{+}0.25] & \textcolor{better}{13} & 5 & \textcolor{worse}{3} & \ensuremath{+}0.65 & \textbf{.023} & \textbf{.007} \\
\midrule
\multicolumn{12}{l}{\textit{Correctness}} \\
Visible tests passed (\%) $\uparrow$ & 29 & 51.1 & 50.1 & \downbad{1.0} & [\ensuremath{-}6.3, \ensuremath{+}4.1] & \textcolor{better}{12} & 3 & \textcolor{worse}{14} & \ensuremath{-}0.04 & .861 & .705 \\
All tests passed (\%) $\uparrow$ & 29 & 40.3 & 40.4 & \upgood{0.1} & [\ensuremath{-}4.8, \ensuremath{+}5.4] & \textcolor{better}{12} & 2 & \textcolor{worse}{15} & \ensuremath{-}0.04 & .878 & .999 \\
\midrule
\multicolumn{12}{l}{\textit{Architecture}} \\
Symbol placement accuracy $\uparrow$ & 27 & 0.95 & 0.92 & \downbad{0.03} & [\ensuremath{-}0.06, \ensuremath{+}0.01] & \textcolor{better}{6} & 8 & \textcolor{worse}{13} & \ensuremath{-}0.43 & .104 & .223 \\
Import-edge precision $\uparrow$ & 29 & 0.70 & 0.69 & \downbad{0.01} & [\ensuremath{-}0.05, \ensuremath{+}0.05] & \textcolor{better}{12} & 0 & \textcolor{worse}{17} & \ensuremath{-}0.17 & .429 & .819 \\
Import-edge recall $\uparrow$ & 29 & 0.58 & 0.61 & \upgood{0.03} & [\ensuremath{-}0.02, \ensuremath{+}0.08] & \textcolor{better}{18} & 1 & \textcolor{worse}{10} & \ensuremath{+}0.39 & .070 & .267 \\
Import-edge F1 $\uparrow$ & 29 & 0.61 & 0.63 & \upgood{0.02} & [\ensuremath{-}0.03, \ensuremath{+}0.06] & \textcolor{better}{18} & 0 & \textcolor{worse}{11} & \ensuremath{+}0.26 & .238 & .559 \\
Modules created & 29 & 18.4 & 19.8 & \ensuremath{+}1.4 & [\ensuremath{-}0.5, \ensuremath{+}3.6] & 18 & 1 & 10 & \ensuremath{+}0.33 & .121 & .106 \\
Lines of code (k) & 29 & 4.59 & 4.58 & \ensuremath{-}0.01 & [\ensuremath{-}0.67, \ensuremath{+}0.48] & 16 & 0 & 13 & \ensuremath{+}0.17 & .429 & .951 \\
\midrule
\multicolumn{12}{l}{\textit{Cost and behaviour}} \\
Wall time (h) $\downarrow$ & 29 & 2.38 & 2.34 & \downgood{0.04} & [\ensuremath{-}0.25, \ensuremath{+}0.18] & \textcolor{better}{16} & 0 & \textcolor{worse}{13} & \ensuremath{-}0.10 & .639 & .658 \\
Held-out pp per 100k output tokens $\uparrow$ & 29 & 13.50 & 14.84 & \upgood{1.34} & [\ensuremath{-}0.80, \ensuremath{+}3.68] & \textcolor{better}{13} & 1 & \textcolor{worse}{15} & \ensuremath{+}0.17 & .438 & .386 \\
Finished by the agent (\%) $\uparrow$ & 29 & 14.9 & 20.7 & \upgood{5.8} & [\ensuremath{-}4.6, \ensuremath{+}17.2] & \textcolor{better}{11} & 10 & \textcolor{worse}{8} & \ensuremath{+}0.20 & .410 & .249 \\
Steps to 50\% visible passing $\downarrow$ & 17 & 297 & 325 & \upbad{28} & [\ensuremath{-}42, \ensuremath{+}108] & \textcolor{better}{9} & 0 & \textcolor{worse}{8} & \ensuremath{+}0.05 & .890 & .502 \\
Steps to 80\% visible passing $\downarrow$ & 13 & 398 & 368 & \downgood{30} & [\ensuremath{-}130, \ensuremath{+}65] & \textcolor{better}{7} & 0 & \textcolor{worse}{6} & \ensuremath{-}0.21 & .542 & .149 \\
Searches for the reference $\downarrow$ & 29 & 5.5 & 4.9 & \downgood{0.6} & [\ensuremath{-}1.7, \ensuremath{+}0.7] & \textcolor{better}{18} & 0 & \textcolor{worse}{11} & \ensuremath{-}0.28 & .194 & .379 \\
\bottomrule
\multicolumn{12}{@{}l}{\footnotesize $^{*}$Without the libraries whose contract names every module.} \\
\multicolumn{12}{@{}l}{\footnotesize $^{\dagger}$Without the libraries whose two implementations share fewer than two classes or functions outside the contract.} \\
\multicolumn{12}{@{}l}{\footnotesize $^{\ddagger}$Without the libraries whose requirements and architecture name no module outside the contract.} \\
\end{tabular}
\end{table*}

\begin{table*}[htbp!]
\centering
\small
\caption{Per-library results, \textbf{\mGemma only}: mean over replicates. \cFull: intent, requirements and C4 architecture; \cIntent: intent only. $^{\dagger}$\textsc{Commit0-Lite}. --: undefined, because the contract names every module of the library.}
\label{tab:agent_libraries_gemma}
\begin{tabular}{lrrrrrrrrrr}
\toprule
 & & \multicolumn{3}{c}{\textbf{Held-out passed (\%)}} & \multicolumn{3}{c}{\textbf{Module F1, non-contract}} & \multicolumn{3}{c}{\textbf{Module F1, all}} \\
\cmidrule(lr){3-5} \cmidrule(lr){6-8} \cmidrule(lr){9-11}
\textbf{Library} & \textbf{Held-out tests} & \textbf{\cIntent} & \textbf{\cFull} & $\Delta$ & \textbf{\cIntent} & \textbf{\cFull} & $\Delta$ & \textbf{\cIntent} & \textbf{\cFull} & $\Delta$ \\
\midrule
attrs & 960 & 0.0 & 48.0 & \upgood{48.0} & 0.81 & 0.90 & \upgood{0.09} & 0.80 & 0.88 & \upgood{0.08} \\
babel$^{\dagger}$ & 4387 & 4.1 & 26.0 & \upgood{21.9} & 0.22 & 0.10 & \downbad{0.12} & 0.50 & 0.50 & 0.00 \\
bitstring & 590 & 9.7 & 7.8 & \downbad{1.9} & 0.20 & 0.26 & \upgood{0.06} & 0.64 & 0.63 & \downbad{0.01} \\
cachetools$^{\dagger}$ & 159 & 52.8 & 47.4 & \downbad{5.4} & -- & -- & -- & 0.69 & 0.79 & \upgood{0.10} \\
chardet$^{\dagger}$ & 376 & 20.8 & 14.4 & \downbad{6.4} & 0.39 & 0.32 & \downbad{0.07} & 0.44 & 0.37 & \downbad{0.07} \\
click & 420 & 25.9 & 17.0 & \downbad{8.9} & 0.77 & 0.84 & \upgood{0.07} & 0.93 & 0.95 & \upgood{0.02} \\
cookiecutter$^{\dagger}$ & 243 & 32.2 & 40.7 & \upgood{8.5} & -- & -- & -- & 0.90 & 1.00 & \upgood{0.10} \\
deprecated$^{\dagger}$ & 120 & 36.1 & 56.1 & \upgood{20.0} & -- & -- & -- & 0.90 & 0.95 & \upgood{0.05} \\
flask & 339 & 1.8 & 25.7 & \upgood{23.9} & 0.56 & 0.53 & \downbad{0.03} & 0.83 & 0.82 & \downbad{0.01} \\
graphene & 293 & 5.5 & 13.9 & \upgood{8.4} & 0.83 & 0.80 & \downbad{0.03} & 0.86 & 0.84 & \downbad{0.02} \\
imapclient$^{\dagger}$ & 192 & 35.4 & 32.3 & \downbad{3.1} & 0.30 & 0.38 & \upgood{0.08} & 0.62 & 0.62 & 0.00 \\
jinja$^{\dagger}$ & 604 & 1.0 & 0.8 & \downbad{0.2} & 0.45 & 0.64 & \upgood{0.19} & 0.57 & 0.61 & \upgood{0.04} \\
loguru & 1046 & 4.6 & 8.9 & \upgood{4.3} & 0.36 & 0.64 & \upgood{0.28} & 0.51 & 0.81 & \upgood{0.30} \\
marshmallow$^{\dagger}$ & 858 & 73.3 & 40.2 & \downbad{33.1} & 0.75 & 0.80 & \upgood{0.05} & 0.90 & 0.93 & \upgood{0.03} \\
more-itertools & 464 & 73.3 & 21.6 & \downbad{51.7} & -- & -- & -- & 1.00 & 0.93 & \downbad{0.07} \\
moviepy & 74 & 47.3 & 45.9 & \downbad{1.4} & 0.36 & 0.61 & \upgood{0.25} & 0.53 & 0.84 & \upgood{0.31} \\
parsel$^{\dagger}$ & 145 & 26.2 & 47.1 & \upgood{20.9} & -- & -- & -- & 0.90 & 1.00 & \upgood{0.10} \\
pexpect & 180 & 45.3 & 40.7 & \downbad{4.6} & 0.63 & 0.62 & \downbad{0.01} & 0.84 & 0.77 & \downbad{0.07} \\
portalocker$^{\dagger}$ & 18 & 27.8 & 18.5 & \downbad{9.3} & 0.61 & 0.79 & \upgood{0.18} & 0.88 & 0.93 & \upgood{0.05} \\
pyjwt$^{\dagger}$ & 182 & 75.1 & 73.3 & \downbad{1.8} & 0.58 & 0.86 & \upgood{0.28} & 0.89 & 0.96 & \upgood{0.07} \\
python-progressbar & 257 & 21.0 & 22.4 & \upgood{1.4} & 0.21 & 0.34 & \upgood{0.13} & 0.62 & 0.68 & \upgood{0.06} \\
python-rsa & 62 & 65.1 & 71.5 & \upgood{6.4} & 0.67 & 0.83 & \upgood{0.16} & 0.99 & 0.98 & \downbad{0.01} \\
requests & 413 & 88.2 & 84.7 & \downbad{3.5} & 0.95 & 0.82 & \downbad{0.13} & 0.99 & 0.96 & \downbad{0.03} \\
simpy$^{\dagger}$ & 101 & 4.3 & 45.9 & \upgood{41.6} & 0.84 & 0.62 & \downbad{0.22} & 0.92 & 0.76 & \downbad{0.16} \\
sqlparse & 287 & 19.2 & 13.2 & \downbad{6.0} & 0.64 & 0.49 & \downbad{0.15} & 0.78 & 0.71 & \downbad{0.07} \\
tinydb$^{\dagger}$ & 144 & 53.5 & 57.6 & \upgood{4.1} & 0.50 & 0.77 & \upgood{0.27} & 0.89 & 0.95 & \upgood{0.06} \\
virtualenv & 217 & 13.6 & 23.8 & \upgood{10.2} & 0.52 & 0.54 & \upgood{0.02} & 0.80 & 0.80 & 0.00 \\
voluptuous$^{\dagger}$ & 104 & 34.6 & 43.9 & \upgood{9.3} & 0.72 & 1.00 & \upgood{0.28} & 0.89 & 1.00 & \upgood{0.11} \\
wcwidth$^{\dagger}$ & 28 & 47.6 & 70.2 & \upgood{22.6} & 0.67 & 0.89 & \upgood{0.22} & 0.61 & 0.89 & \upgood{0.28} \\
\midrule
Mean &  & 32.6 & 36.5 & \upgood{3.9} & 0.56 & 0.64 & \upgood{\textbf{0.08}} & 0.78 & 0.82 & \upgood{0.04} \\
\bottomrule
\end{tabular}
\end{table*}

\begin{table*}[htbp!]
\centering
\small
\caption{Per-library results, \textbf{\mQwen only}: mean over replicates. \cFull: intent, requirements and C4 architecture; \cIntent: intent only. $^{\dagger}$\textsc{Commit0-Lite}. --: undefined, because the contract names every module of the library.}
\label{tab:agent_libraries_qwen}
\begin{tabular}{lrrrrrrrrrr}
\toprule
 & & \multicolumn{3}{c}{\textbf{Held-out passed (\%)}} & \multicolumn{3}{c}{\textbf{Module F1, non-contract}} & \multicolumn{3}{c}{\textbf{Module F1, all}} \\
\cmidrule(lr){3-5} \cmidrule(lr){6-8} \cmidrule(lr){9-11}
\textbf{Library} & \textbf{Held-out tests} & \textbf{\cIntent} & \textbf{\cFull} & $\Delta$ & \textbf{\cIntent} & \textbf{\cFull} & $\Delta$ & \textbf{\cIntent} & \textbf{\cFull} & $\Delta$ \\
\midrule
attrs & 960 & 0.0 & 0.0 & 0.0 & 0.73 & 0.90 & \upgood{0.17} & 0.79 & 0.83 & \upgood{0.04} \\
babel$^{\dagger}$ & 4387 & 7.4 & 13.7 & \upgood{6.3} & 0.11 & 0.17 & \upgood{0.06} & 0.31 & 0.53 & \upgood{0.22} \\
bitstring & 590 & 7.1 & 14.9 & \upgood{7.8} & 0.22 & 0.43 & \upgood{0.21} & 0.62 & 0.73 & \upgood{0.11} \\
cachetools$^{\dagger}$ & 159 & 49.3 & 54.9 & \upgood{5.6} & -- & -- & -- & 0.49 & 0.69 & \upgood{0.20} \\
chardet$^{\dagger}$ & 376 & 12.1 & 58.3 & \upgood{46.2} & 0.50 & 0.54 & \upgood{0.04} & 0.54 & 0.58 & \upgood{0.04} \\
click & 420 & 32.1 & 28.4 & \downbad{3.7} & 0.86 & 0.89 & \upgood{0.03} & 0.96 & 0.97 & \upgood{0.01} \\
cookiecutter$^{\dagger}$ & 243 & 38.0 & 25.5 & \downbad{12.5} & -- & -- & -- & 0.99 & 0.90 & \downbad{0.09} \\
deprecated$^{\dagger}$ & 120 & 37.2 & 35.8 & \downbad{1.4} & -- & -- & -- & 0.82 & 0.90 & \upgood{0.08} \\
flask & 339 & 38.2 & 24.5 & \downbad{13.7} & 0.55 & 0.39 & \downbad{0.16} & 0.83 & 0.57 & \downbad{0.26} \\
graphene & 293 & 9.6 & 7.2 & \downbad{2.4} & 0.61 & 0.79 & \upgood{0.18} & 0.67 & 0.84 & \upgood{0.17} \\
imapclient$^{\dagger}$ & 192 & 37.3 & 43.1 & \upgood{5.8} & 0.29 & 0.63 & \upgood{0.34} & 0.83 & 0.93 & \upgood{0.10} \\
jinja$^{\dagger}$ & 604 & 0.0 & 1.5 & \upgood{1.5} & 0.28 & 0.29 & \upgood{0.01} & 0.29 & 0.31 & \upgood{0.02} \\
loguru & 1046 & 4.2 & 3.8 & \downbad{0.4} & 0.24 & 0.41 & \upgood{0.17} & 0.43 & 0.54 & \upgood{0.11} \\
marshmallow$^{\dagger}$ & 858 & 76.3 & 65.6 & \downbad{10.7} & 0.72 & 0.96 & \upgood{0.24} & 0.90 & 0.99 & \upgood{0.09} \\
more-itertools & 464 & 48.6 & 59.8 & \upgood{11.2} & -- & -- & -- & 0.67 & 0.92 & \upgood{0.25} \\
moviepy & 74 & 49.5 & 50.9 & \upgood{1.4} & 0.54 & 0.66 & \upgood{0.12} & 0.81 & 0.84 & \upgood{0.03} \\
parsel$^{\dagger}$ & 145 & 31.3 & 13.1 & \downbad{18.2} & -- & -- & -- & 1.00 & 0.96 & \downbad{0.04} \\
pexpect & 180 & 28.9 & 24.3 & \downbad{4.6} & 0.65 & 0.68 & \upgood{0.03} & 0.86 & 0.84 & \downbad{0.02} \\
portalocker$^{\dagger}$ & 18 & 46.3 & 37.0 & \downbad{9.3} & 0.27 & 0.67 & \upgood{0.40} & 0.77 & 0.89 & \upgood{0.12} \\
pyjwt$^{\dagger}$ & 182 & 74.9 & 76.4 & \upgood{1.5} & 0.58 & 0.86 & \upgood{0.28} & 0.89 & 0.96 & \upgood{0.07} \\
python-progressbar & 257 & 17.0 & 13.2 & \downbad{3.8} & 0.17 & 0.29 & \upgood{0.12} & 0.63 & 0.69 & \upgood{0.06} \\
python-rsa & 62 & 68.8 & 67.2 & \downbad{1.6} & 0.00 & 0.33 & \upgood{0.33} & 0.95 & 0.95 & 0.00 \\
requests & 413 & 88.1 & 54.2 & \downbad{33.9} & 0.88 & 0.88 & 0.00 & 0.97 & 0.97 & 0.00 \\
simpy$^{\dagger}$ & 101 & 34.0 & 40.9 & \upgood{6.9} & 0.83 & 0.91 & \upgood{0.08} & 0.92 & 0.96 & \upgood{0.04} \\
sqlparse & 287 & 16.0 & 36.4 & \upgood{20.4} & 0.73 & 0.79 & \upgood{0.06} & 0.84 & 0.88 & \upgood{0.04} \\
tinydb$^{\dagger}$ & 144 & 44.0 & 60.4 & \upgood{16.4} & 0.27 & 0.72 & \upgood{0.45} & 0.82 & 0.93 & \upgood{0.11} \\
virtualenv & 217 & 27.6 & 24.4 & \downbad{3.2} & 0.45 & 0.50 & \upgood{0.05} & 0.68 & 0.76 & \upgood{0.08} \\
voluptuous$^{\dagger}$ & 104 & 34.3 & 26.6 & \downbad{7.7} & 0.77 & 0.67 & \downbad{0.10} & 0.92 & 0.89 & \downbad{0.03} \\
wcwidth$^{\dagger}$ & 28 & 59.5 & 48.8 & \downbad{10.7} & 0.96 & 0.67 & \downbad{0.29} & 0.91 & 0.61 & \downbad{0.30} \\
\midrule
Mean &  & 35.1 & 34.9 & \downbad{0.2} & 0.51 & 0.63 & \upgood{\textbf{0.12}} & 0.76 & 0.80 & \upgood{\textbf{0.04}} \\
\bottomrule
\end{tabular}
\end{table*}

\fi 

\end{document}